\documentclass[aps,prd,twocolumn,superscriptaddress,amsmath,amssymb,nofootinbib,longbibliography,preprintnumbers,floatfix]{revtex4-2}
\usepackage[dvipsnames]{xcolor}
\usepackage[colorlinks]{hyperref}
\usepackage{xspace}

\usepackage{tikz}
\usetikzlibrary{tikzmark}
\usetikzlibrary{positioning}
\usepackage[normalem]{ulem}

\hypersetup{linkcolor=BrickRed,citecolor=Green,
filecolor=Mulberry,
urlcolor=NavyBlue,
menucolor=BrickRed,
runcolor=Mulberry
}

\definecolor{lime}{HTML}{A6CE39}
\DeclareRobustCommand{\orcidicon}{\hspace{-1mm}
	\begin{tikzpicture}
	\draw[lime, fill=lime] (0,0) 
	circle [radius=0.16] 
	node[white] {{\fontfamily{qag}\selectfont \tiny \,ID}};
	\draw[white, fill=white] (-0.0525,0.095) 
	circle [radius=0.007];
	\end{tikzpicture}
	\hspace{-3mm}
}

\foreach \x in {A, ..., Z}{\expandafter\xdef\csname orcid\x\endcsname{\noexpand\href{https://orcid.org/\csname orcidauthor\x\endcsname}
			{\noexpand\orcidicon}}
}

\usepackage{graphicx}
\usepackage{bbm}

\makeatletter
\newcommand{\abs}{\@ifstar\abssmall\absbig}
\newcommand{\absbig}[1]{\left \lvert #1 \right \rvert}
\newcommand{\abssmall}[1]{\lvert #1 \rvert}
\makeatother

\newcommand{\vk}{\vec{k}}
\newcommand{\x}{\mathrm{x}}
\newcommand{\y}{\mathrm{y}}
\newcommand{\z}{\mathrm{z}}
\renewcommand{\i}{\mathrm{i}}
\newcommand{\Tr}{\mathrm{Tr}}
\renewcommand{\Im}{\mathrm{Im}}

\newcommand{\pdv}[2]{\frac{\partial #1}{\partial #2}}

\newcommand{\dd}{\mathrm{d}}

\newcommand{\bN}{\overline{N}}
\newcommand{\bF}{\overline{F}}
\newcommand{\bP}{\overline{P}}
\newcommand{\bA}{\bar{A}}
\newcommand{\bB}{\overline{B}}

\newcommand{\bPi}{\overline{\Pi}}
\newcommand{\bDelta}{\overline{\Delta}}

\newcommand{\nbnuin}{\mathrel{\smash{\overset{\makebox[0pt]{\mbox{\tiny (---)}}}{\nu}}}}

\newcommand{\nbDelta}{\mathrel{\overset{\makebox[0pt]{\mbox{\tiny (---)}}}{\Delta}}}

\newcommand{\nbN}{\mathrel{\overset{\makebox[0pt]{\mbox{\tiny (---)}}}{N}}}
\newcommand{\nbNin}{\mathrel{\smash{\overset{\makebox[0pt]{\mbox{\tiny (---)}}}{N}}}}
\newcommand{\nbF}{\mathrel{\overset{\makebox[0pt]{\mbox{\tiny (---)}}}{F}}}

\newcommand{\vrho}{\varrho}
\newcommand{\bvrho}{\bar{\varrho}}
\newcommand{\N}{\mathcal{N}}
\newcommand{\F}{\mathcal{F}}
\renewcommand{\P}{\mathcal{P}}
\renewcommand{\S}{\mathcal{S}}
\newcommand{\M}{\mathcal{M}}

\newcommand{\veps}{\varepsilon}

\newcommand{\flash}{\texttt{FLASH}\xspace}
\newcommand{\emu}{\texttt{Emu}\xspace}

\renewcommand{\vec}{\mathbf}

\begin{document}
\preprint{N3AS-23-032}

\title{Neutrino fast flavor oscillations with moments: linear stability analysis and application to neutron star mergers}

\author{Julien Froustey\orcidA{}}
\email{jfroustey@berkeley.edu}
\affiliation{Department of Physics, North Carolina State University, Raleigh, NC 27695, USA}
\affiliation{Department of Physics, University of California Berkeley, Berkeley, CA 94720, USA}

\author{Sherwood Richers\orcidC{}}
\affiliation{Department of Physics, University of Tennessee Knoxville, Knoxville, TN 37996, USA}

\author{Evan Grohs\orcidD{}}
\affiliation{Department of Physics, North Carolina State University, Raleigh, NC 27695, USA}

\author{Samuel D. Flynn}
\affiliation{Department of Physics, North Carolina State University, Raleigh, NC 27695, USA}

\author{Francois Foucart\orcidF{}}
\affiliation{Department of Physics \& Astronomy, University of New Hampshire, 9 Library Way, Durham, NH 03824, USA}

\author{James P. Kneller\orcidG{}}
\affiliation{Department of Physics, North Carolina State University, Raleigh, NC 27695, USA}

\author{Gail C. McLaughlin\orcidH{}}
\affiliation{Department of Physics, North Carolina State University, Raleigh, NC 27695, USA}

\begin{abstract}

Providing an accurate modeling of neutrino physics in dense astrophysical environments such as binary neutron star mergers presents a challenge for hydrodynamic simulations. Nevertheless, understanding how flavor transformation can occur and affect the dynamics, the mass ejection, and the nucleosynthesis will need to be achieved in the future. Computationally expensive, large-scale simulations frequently evolve the first classical angular moments of the neutrino distributions. By promoting these quantities to matrices in flavor space, we develop a linear stability analysis of fast flavor oscillations using only the first two “quantum” moments, which notably requires generalizing the classical closure relations that appropriately truncate the hierarchy of moment equations in order to treat quantum flavor coherence. After showing the efficiency of this method on a well-understood test situation, we perform a systematic search of the occurrence of fast flavor instabilities in a neutron star merger simulation. We discuss the successes and shortcomings of moment linear stability analysis, as this framework provides a time-efficient way to design and study better closure prescriptions in the future.
\end{abstract}

\maketitle

\section{Introduction}

A thorough understanding of neutrino physics in dense astrophysical environments is crucial if we are to fully comprehend systems such as core-collapse supernovae (CCSNe) or neutron star mergers (NSMs). Neutrino emission and transport impact the explosion mechanism of CCSNe, mass ejection from NSMs, and the subsequent nucleosynthesis in both. While multiple advances have been made in the level of detail of the physics that goes into state-of-the-art simulations (radiation transport, general relativistic corrections, magnetohydrodynamics, equation of state, progenitor modeling, etc.)~\cite{Etienne:2011ea,Kiuchi:2014hja,Wanajo:2014wha,Neilsen:2014hha,Perego:2015agy,Foucart:2015gaa,Foucart:2016rxm,Sekiguchi:2016bjd,Ardevol-Pulpillo:2018btx,Gizzi:2019awu,Foucart:2021mcb,Radice:2021jtw,Cook:2023bag}, neutrino flavor oscillations — which might play an important role — are not yet included consistently. 

As with many other aspects of CCSNe and NSMs, the extreme conditions in these environments mean that the neutrino flavor oscillations are far from simple. A neutrino will encounter multiple regimes in such systems~\cite{Volpe_review} and the flavor transformation will be driven by different contributions from the mean-field Hamiltonian which describes flavor evolution. The flavor transformation regimes include MSW resonances \cite{Wolfenstein:1977ue,Mikheev:1986wj,Schirato:2002tg,Fogli:2003dw,Choubey:2006aq,Kneller:2007kg}, parametric resonances due to turbulence \cite{Loreti:1995ae,Kneller:2010sc,Borriello:2013tha,Patton:2014lza,Yang:2015oya}, matter-neutrino resonances~\cite{Malkus:2012ts,Malkus:2015mda,Vaananen:2015hfa,Wu:2015fga,Zhu:2016mwa,Shalgar:2017pzd,Vlasenko:2018irq}, and “slow” collective oscillations~\cite{Duan_review,2009PhRvL.103g1101G,Galais:2011jh}. Recently, so-called “fast” flavor instabilities (FFIs), which are characterized by a typical growth rate for the flavor coherence of the order of $\sim \sqrt{2} G_F n_\nu$ with $G_F$ Fermi's constant and $n_\nu$ the neutrino density, have received a lot of attention (see \cite{Tamborra_review,Capozzi_review,Richers_review} for recent reviews). In a two-flavor system, these instabilities are known to be triggered by the presence of an angular crossing in the difference of electron lepton number (ELN) and heavy-lepton flavor number (XLN) distributions. In other words, at a given location there must be directions in which the lepton number difference current has opposite signs, meaning that there is a direction in which the lepton number difference current crosses from positive to negative. 

Initially discovered with toy models~\cite{Sawyer:2005jk,Sawyer:2008zs}, FFIs were subsequently studied in dense astrophysical environments. Scans of CCSN and NSM simulations which employed discrete ordinate transport (but which did not include oscillations) indicate the conditions that lead to the FFI can be found throughout the simulation~\cite{Abbar:2018shq,Abbar:2019zoq,Nagakura:2019sig,DelfanAzari:2019tez,Harada:2021ata}. In order to determine if the FFI should occur in classical simulations which use \emph{moment} transport, one must undertake an angular reconstruction, of which many variations have appeared in the literature (e.g., \cite{Richers:2022dqa,Morinaga:2019wsv,Abbar:2020qpi,Capozzi:2020syn,Nagakura:2021hyb,Nagakura:2021txn,Nagakura:2021suv}). The FFI has been studied through both linear stability analysis (e.g.,~\cite{Izaguirre:2016gsx,Dasgupta:2016dbv,Capozzi:2017gqd,Morinaga:2018aug,DelfanAzari:2019epo,Abbar:2021lmm,Padilla-Gay:2021haz}) and numerical calculations of small regions (see~\cite{Richers:2022bkd} and references therein), along with analytical works on the structure and solutions of the equations of motion~\cite{Fiorillo:2023mze,Fiorillo:2023hlk}. In addition, various groups investigated the effect of collisions on fast flavor conversions~\cite{Shalgar:2020wcx,Martin:2021xyl,Sigl:2021tmj,Sasaki:2021zld,Hansen:2022xza}, which also led to the discovery of a new kind of instability, called the collisional instability~\cite{Johns:2021qby,Padilla-Gay:2022wck,Johns:2022yqy,Lin:2022dek,Xiong:2022vsy,Xiong:2022zqz,Liu:2023pjw,Liu:2023vtz,Akaho:2023brj,Fiorillo:2023ajs}.

Although the conditions found in CCSNe and NSMs are such that the lengthscale ($\sim \mathrm{cm}$) and timescale ($\sim 0.1 \, \mathrm{ns}$) of the FFI currently prevent our ability to solve for the flavor transformation concurrent with the hydrodynamics, several recent studies have nevertheless aimed to assess the potential effect of flavor transformation using prescriptions that inexpensively approximate this subgrid physics~\cite{Just:2022flt,Li:2021vqj,Ehring:2023lcd,Ehring:2023abs}. A second, alternative approach to the huge problem of disparate length- and time- scales consists of artificially reducing the strength of the neutrino self-interaction potential responsible for fast flavor oscillations in order to increase the length- and time- scales of the oscillations to the point where FFI can be included in a large-scale simulation. The results are then extrapolated to a realistic interaction strength~\cite{Nagakura:2022kic,Nagakura:2022xwe,Nagakura:2023mhr,Nagakura:2023wbf}.

Finally a third approach is to generalize the classical angular moments to quantum moments \cite{strack:2005,Zhang:2013lka,Myers:2021hnp} and evolve a limited number of them (e.g., the number densities and fluxes) rather than a much larger number of ordinates. 
The quantum kinetic equations for neutrinos traveling in different directions become a hierarchy of evolution equations for the moments which one truncates using a closure relation (e.g., \cite{Shibata:2011kx,Zhang:2013lka,Richers:2019grc}, though the truncated moment formalism in general dates back to~\cite{israel1979transient, thorne1981relativistic}). Such a method was used in~\cite{Myers:2021hnp} to describe neutrino flavor transformation in a spherically symmetric bulb model, and it was subsequently shown in~\cite{Grohs:2022fyq} that much of the phenomenology of the FFI can be captured directly by quantum moments. 
Given the reduced number of quantities that are numerically evolved in a quantum moment method plus the straightforward connection with classical moment transport, the computational efficiency of quantum moment transport is very attractive. However, the timescales associated with FFIs are so small that even a moment code cannot be used to describe the transport everywhere in large-scale simulations. A more feasible approach may be to use it only where quantum effects are important. The key to identifying where and when to use a quantum moment approach is to understand better where the FFI can take place. To that end, linear stability analysis (LSA) plays a crucial role in determining the time- and length- scales of FFI, while requiring considerably less computation time. 
Linear stability analysis provides a powerful framework to efficiently determine the possibility and the growth characteristics of fast flavor conversion in virtually any scenario. However, it cannot say anything about the quasi-steady state reached after the saturation of the instability. Notably, the growth rate is not related to the final amount of flavor conversion in general~\cite{Padilla-Gay:2021haz,Nagakura:2022xwe}. Notwithstanding these shortcomings, LSA could act as a useful guide for determining when a quantum moment treatment is necessary. The power of LSA has long been recognized and previous works have designed such a moment LSA but necessarily restricted the analysis to the “zero mode”, which is the homogeneous mode in a particular co-rotating frame~\cite{Dasgupta:2018ulw,Glas:2019ijo}.

In this paper, we introduce a generic two-moment linear stability analysis for arbitrary wavenumbers, that is able to predict successfully the presence and main characteristics of the growth phase of FFI in a variety of situations. After demonstrating that it works in several test problems, we apply our method to a snapshot of a NSM simulation~\cite{Foucart:2015gaa,Foucart:2016rxm}, in order to perform a systematic and time-efficient prediction of the existence of FFI in various regions post-merger. We then compare our LSA predictions with numerical simulations of the neutrino evolution using both a moment method which follows closely the assumptions of our LSA (\flash~\cite{Grohs:2022fyq,Grohs:2023pgq}) and with a particle-in-cell multi-angle code (\emu~\cite{Richers_emu}).

Our paper is organized as follows. In Sec.~\ref{sec:QKE}, we introduce the quantum kinetic equations adapted to the study of FFI along with their moment formulation (see also Appendix~\ref{app:moments}). The linear stability analysis is presented in Sec.~\ref{sec:LSA}. We apply this moment LSA to a test case and compare it to numerical calculations of the full QKEs in Sec.~\ref{sec:fiducial} before focusing on the NSM simulation in Sec.~\ref{sec:NSM}. In Sec.~\ref{sec:conclusion} we give conclusions and discuss how we may refine the current LSA for future applications. Technical details, regarding the self-consistency of our results, are gathered in Appendix~\ref{app:multi_angle_LSA}, and for completeness, we assess the regions where collisions might play a role through collisional instabilities in Appendix~\ref{app:collisional_inst}.

Throughout this paper, we work in natural units where $\hbar=c=1$.

\section{Quantum Kinetic Equations}
\label{sec:QKE}

In astrophysical and cosmological environments, neutrino transport which takes into account flavor mixing uses the formalism of quantum kinetic equations (QKEs)~\cite{SiglRaffelt,Vlasenko:2013fja,Blaschke:2016xxt,Froustey:2020mcq}. These equations describe the evolution of the one-body reduced density matrices $\vrho(t,\vec{x},\vec{p})$ and $\bvrho(t,\vec{x},\vec{p})$ for neutrinos and antineutrinos. They are, in general, $2 n_f \times 2 n_f$ Hermitian matrices for $n_f$ flavors and $2$ chiral states. In this work, with the typical energies being far larger than the neutrino masses, we only consider left-handed neutrinos and right-handed antineutrinos and neglect spin coherence terms~\cite{Cirigliano:2014aoa}. The density matrices are thus $n_f \times n_f$ matrices, whose diagonal entries correspond to the classical distribution functions, while the complex off-diagonal components account for flavor coherence.

We will only consider two-flavor mixing, between the electronic (anti)neutrino state and the “heavy lepton” flavor state $x$, so that the density matrices read
\begin{equation}
\vrho(t, \vec{x}, \vec{p}) = \begin{pmatrix} \vrho_{e e} & \vrho_{e x} \\ \vrho_{x e} & \vrho_{x x} \end{pmatrix} \ , \ \ \bvrho(t, \vec{x}, \vec{p}) = \begin{pmatrix} \bvrho_{e e} & \bvrho_{e x} \\ \bvrho_{x e} & \bvrho_{x x} \end{pmatrix} \, .
\end{equation}

The QKE for $\vrho$ is usually written:
\begin{equation}
\label{eq:QKE}
\i \left( \pdv{\vrho_{ab}}{t} + \dot{\vec{x}} \cdot \vec{\nabla} \vrho_{ab} \right) = \left[H, \vrho\right]_{ab} + \i \,  C_{ab} \, ,
\end{equation}
with $C_{ab}$ the collision term and where a dot denotes a time derivative. $H$ is the Hamiltonian-like operator which generally describes the kinetic energy (vacuum oscillations), matter and self-interaction mean-fields. Fast flavor oscillations, which are the focus of this paper, are driven by the self-interaction Hamiltonian:
\begin{equation}
\label{eq:H_self}
H_{\nu \nu}(\vec{p}) = \frac{\sqrt{2} G_F}{(2 \pi)^3} \! \int{\! \vec{d q}(1-\cos \vartheta)[\vrho(t,\vec{x},\vec{q}) - \bvrho^*(t,\vec{x},\vec{q})]} \, ,
\end{equation}
with $\vartheta$ the angle between $\vec{p}$ and $\vec{q}$. There is a similar equation for $\bvrho$, with the Hamiltonian $\bar{H}_{\nu\nu}=-H_{\nu \nu}^*$.

The typical timescale of fast flavor instabilities is given by $\Delta t^{-1} \sim \sqrt{2} G_F n_\nu$ (with $n_\nu$ the neutrino density), which can be much shorter than the vacuum oscillation timescale or the inverse collision rate. Because of this, we consider a simplified QKE where only the self-interaction mean-field is kept on the right-hand side. This means we do not capture a multitude of other flavor transformation phenomena, although we briefly discuss collisional instabilities in Appendix~\ref{app:collisional_inst} and plan to address them in more detail in future work. From now on we simplify the system of equations by considering mono-energetic (anti)neutrinos with common energy $p$, although this approximation should be relaxed in the future (see, e.g.,~\cite{Shalgar:2020xns}). 

Under these assumptions, the QKEs for $\vrho$ and $\bvrho$ read:
\begin{align}
\label{eq:QKE_rho}
\i \left( \pdv{\vrho_{ab}}{t} + \dot{\vec{x}} \cdot \vec{\nabla} \vrho_{ab} \right) &= \left[H_{\nu \nu}, \vrho\right]_{ab} \, , \\
\label{eq:QKE_rhobar}
\i \left( \pdv{\bvrho_{ab}}{t} + \dot{\vec{x}} \cdot \vec{\nabla} \bvrho_{ab} \right) &= - \left[H_{\nu \nu}^*, \bvrho\right]_{ab} \, .
\end{align}

\subsection{Moment equations}

The first angular moments of the neutrino distribution are the number density, number flux and pressure tensor:\footnote{Some references define these moments per unit solid angle (in which case the expressions~\eqref{eq:moments} must be divided by $4 \pi$), as in~\cite{Myers:2021hnp}. We chose the same convention as~\cite{Grohs:2022fyq,Grohs:2023pgq}, in order to make equations more readable. Note also that we are using the term “pressure” for the second moment even though our definition of this moment does not have the actual units of a pressure (a better name may be “number pressure”).}
\begin{subequations}
\label{eq:moments}
\begin{align}
    N_{ab}(t,\vec{x}) &\equiv \int{\dd p} \, \frac{p^2}{(2 \pi)^3} \int{\dd \vec{\Omega} \, \vrho_{ab}(t,\vec{x},\vec{p})} \, , \\
    F_{ab}^i(t,\vec{x}) &\equiv \int{\dd p} \,  \frac{p^2}{(2 \pi)^3}  \int{\dd \vec{\Omega} \, \frac{p^i}{p} \, \vrho_{ab}(t,\vec{x},\vec{p})} \, , \\
    P_{ab}^{ij}(t,\vec{x}) &\equiv \int{\dd p} \, \frac{p^2}{(2 \pi)^3}  \int{\dd \vec{\Omega} \, \frac{p^i p^j}{p^2} \, \vrho_{ab}(t,\vec{x},\vec{p})} \, .
\end{align}
\end{subequations}
For the mono-energetic system we consider, the integral over the momentum magnitude becomes $\Delta p \times p^2/(2 \pi)^3$, with $\Delta p$ the width of the single energy bin of the system.

Integrating the QKE~\eqref{eq:QKE_rho} over the solid angle of $\vec{p}$, multiplied by $1$ or $p^j/p$, leads to the first two “moment QKEs” (a derivation using a covariant formalism is proposed in the Appendix~\ref{app:moments}). They read, using Einstein's summation convention:
\begin{subequations}
\begin{align}
\i \left(\pdv{N}{t} + \pdv{F^j}{x^j} \right) &= \sqrt{2} G_F \left[N - \bN^*, N\right] \nonumber \\
&\qquad \qquad  - \sqrt{2} G_F  \left[(F-\bF^*)_j,F^j\right] \, ,  \label{eq:QKE_moment_N} \\
\i \left(\pdv{F^i}{t} + \pdv{P^{ij}}{x^j}\right) &= \sqrt{2} G_F \left[N - \bN^*, F^i\right] \nonumber \\
&\qquad \qquad 
- \sqrt{2} G_F \left[(F-\bF^*)_j,P^{ij}\right] \, . \label{eq:QKE_moment_F}
\end{align}
\end{subequations}
The same equations for antineutrino moments are:
\begin{subequations}
\begin{align}
\i \left(\pdv{\bN}{t} + \pdv{\bF^j}{x^j} \right) &= - \sqrt{2} G_F \left[N^* - \bN, \bN \right] \nonumber \\
&\qquad \qquad  + \sqrt{2} G_F  \left[(F^*-\overline{F})_j,\bF^j\right] \, ,  \label{eq:QKE_moment_Nbar} \\
\i \left(\pdv{\bF^i}{t} + \pdv{\bP^{ij}}{x^j}\right) &= - \sqrt{2} G_F \left[N^* - \overline{N}, \bF^i\right] \nonumber \\
&\qquad \qquad 
+ \sqrt{2} G_F \left[(F^*-\overline{F})_j,\bP^{ij}\right] \, . \label{eq:QKE_moment_Fbar}
\end{align}
\end{subequations}
As Eqs.~\eqref{eq:QKE_moment_F} and \eqref{eq:QKE_moment_Fbar} depend on the pressure tensor, a truncated two-moment method requires an appropriate closure, i.e., an expression $P^{ij}(N,\vec{F})$ that is the topic of the next section.

\subsection{Closure}
\label{subsec:closure}

We restrict ourselves to analytic closures for simplicity, although more sophisticated treatments of closures have been used for classical radiation transport (e.g., variable Eddington tensor methods \cite{Hubeny:2006wm,Muller:2010ymw,Jiang:2012yw}). An analytic closure for the truncated moment method specifies the pressure as a function of the number density and flux moments. Various choices are available in the literature (see e.g., \cite{Murchikova:2017zsy}), and we focus here on the options considered in this work and implemented in \flash.

\subsubsection{Classical closure}

In classical transport, the pressure is usually expressed as an interpolation between the optically thin and thick limits as
\begin{equation}
    P^{ij} = \frac{3\chi-1}{2}P^{ij}_{\mathrm{thin}}+\frac{3(1-\chi)}{2}P^{ij}_{\mathrm{thick}}\,\, ,
    \label{eq:Pij_closed}
\end{equation}
where the Eddington factor $\chi$ is generally in the range $1/3 \leq \chi \leq 1$~\cite{Murchikova:2017zsy}. The lower limit $\chi=1/3$ places the pressure in optically thick conditions, and conversely the upper limit $\chi=1$ corresponds to optically thin conditions.
The pressure for these limits reads
\begin{align}
    P^{ij}_\mathrm{thick} &= \frac{N}{3} \delta^{ij} \, , \label{eq:Pthick_class} \\
    P_\mathrm{thin}^{ij} &= N \frac{F^i F^j}{\abs*{\vec{F}}^2} \, .
    \label{eq:Pthin_class}
\end{align}

There are several suggestions of analytic closures in the literature, which are generally expressed as $\chi(\hat{f})$, where $\hat{f} = \abs*{\vec{F}}/N$ is the flux factor. We use here, for consistency with the choice made in the simulation~\cite{Foucart:2015gaa,Foucart:2016rxm} that we shall consider later in this paper, the classical maximum entropy or “Minerbo” closure~\cite{minerbo_maximum_1978,cernohorsky_bludman,Smit_closure,Murchikova:2017zsy}, for which the function $\chi(\hat{f})$ is given to 2 \% accuracy by the approximate polynomial expression:
\begin{equation}
\label{eq:chi_MEC}
\chi(\hat{f}) = \frac{1}{3} + \frac{2 \hat{f}^2}{15}\left(3-\hat{f}+3 \hat{f}^2\right)\, .
\end{equation}

\subsubsection{Quantum closure}
Generalizing to quantum kinetics, the neutrino angular moments are flavor matrices, which makes the closure for the off-diagonal entries a priori non-obvious. Indeed, we cannot interpret for instance $N_{ex}$ as the number density of a given neutrino species, notably because $N_{ex}$ is not constrained to be a positive real number. A detailed study of a proper “quantum” closure is beyond the scope of this work, but it remains a challenge to be addressed in the future to design successful and computationally efficient moment calculations~\cite{kneller23}. In this work, we focus on the simplest generalization of the classical case, hence building a “semiclassical” closure. 

The optically thick limit of the pressure tensor is straightforward to generalize from~\eqref{eq:Pthick_class} and can be expressed by simply elevating the number density to a complex matrix:
\begin{equation}
    \label{eq:Pij_thick}
    P^{ij}_{ab,\mathrm{thick}} = \frac{N_{ab}}{3}\delta^{ij}\,\,.
\end{equation}
Even though this relation is well-motivated for the flavor diagonal components ($a = b$), there is a priori no reason for it to be also true when $a \neq b$ (for instance, the complex flavor off-diagonal elements of $P$ and $N$ could have different phases). The only constraint that must be satisfied for the flavor both on- and off-diagonal components of $P$ is the geometrical relation:
\begin{equation}
\label{eq:trace_P}
    P_{ab}^{ij} \, \delta_{ij} = P_{ab}^{\x \x} + P_{ab}^{\y \y} + P_{ab}^{\z \z} = N_{ab} \, .
\end{equation}
Nevertheless, the positive results that have been obtained with the prescription~\eqref{eq:Pij_thick} (see \cite{Grohs:2022fyq} and Secs.~\ref{sec:fiducial} and \ref{sec:NSM}) warrant using it as a starting point.

Proposing a relationship between $N$ and $P$ in the optically thin (free streaming) limit is more challenging, made even more so when we consider cases where the different flavors (and flavor mixtures) can have fluxes in different directions. One possibility is to define a free streaming limit separately for each component of the flavor matrix. This “component-wise” (CW) version of the free streaming limit would read as:
\begin{equation}
\label{eq:Pij_thin_1}
^{(\mathrm{CW})}P^{ij}_{ab,\mathrm{thin}} = N_{ab} \frac{F_{ab}^i F_{ab}^j}{F^k_{ab}F^l_{ab}\delta_{kl}} = N_{ab} \frac{F_{ab}^i F_{ab}^j}{\abs*{\vec{F}_{ab}}^2}\, .
\end{equation}
The component-wise version allows different flavors to propagate in different directions, but under a change of (flavor) basis, the tensor $P^{ij}_{ab}$ does not undergo the same unitary transformation as $N_{ab}$ or $\vec{F}_{ab}$.

Another option is to use a single representative flux direction for all the flavor components. That is, we “flavor-trace” (FT) the flux part:
\begin{align}
^{(\mathrm{FT})}P^{ij}_{ab,\mathrm{thin}} &= N_{ab} \frac{\left(F_{cd}^i\Delta^{cd}\right)\left( F_{ef}^j\Delta^{ef}\right)}{\left(F_{cd}^k\Delta^{cd}\right)\left( F_{ef}^l\Delta^{ef}\right)\delta_{kl}} \nonumber \\ &= N_{ab} \frac{\Tr(F^i) \Tr(F^j)}{\abs*{\vec{\Tr(F)}^2}} \, , \label{eq:Pij_thin_2}
\end{align}
where we used different notations for the Kronecker delta in space ($\delta_{ij}$) and in flavor space ($\Delta_{ab}$). This version of the free streaming limit has a straightforward interpretation, but does not allow different flavors to have different pressure tensor shapes; they are all tied to the flavor-averaged propagation direction although each component still satisfies Eq.~\eqref{eq:trace_P} individually. 

Once we have decided how to define the free streaming limit, we elevate the interpolant $\chi$ to a flavor matrix $X$ so that the pressure tensor is found by interpolating between its two limits as: 
\begin{equation}
    \label{eq:Pij_closed_quantum}
    P_{ab}^{ij} = \frac{3 X_{ab}-1}{2} P_{ab,\mathrm{thin}}^{ij} + \frac{3(1-X_{ab})}{2} P_{ab,\mathrm{thick}}^{ij} \, .
\end{equation}
As with the pressure in the free streaming limit, we have to decide how to define $X$. One option would be to evaluate the closure separately for each component as
\begin{equation}
\label{eq:chi_ab_nonFT}
^{(\mathrm{CW})}X_{ab} = \chi\left(\frac{\abs*{\vec{F}_{ab}}}{N_{ab}}\right)\,\,.
\end{equation}
This is, once again, not independent of the choice of (flavor) basis. A second option would be to choose flavor-traced quantities as
\begin{equation}
\label{eq:chi_ab}
^{(\mathrm{FT})}X_{ab} = \chi \!\left(\frac{\sqrt{(F_{cd}^i \Delta^{cd})(F_{ef}^j \Delta^{ef})\delta_{ij}}}{N_{cd}\Delta^{cd}}\right) \! = \chi \! \left( \frac{\abs*{\vec{\Tr(F)}}}{\Tr(N)}  \right) ,
\end{equation}
such that all the components of the matrix $^{(\mathrm{FT})}X$ are equal. This nevertheless prevents different flavors from being in different states of free-streaming. 

From this brief discussion the reader may now appreciate that the issue of what constitutes a “quantum closure” is fraught with peril. It is not our intention to explore this issue further in this paper. Our choice for the quantum closure is to adopt the definitions $^{(\mathrm{CW})} P_{ab,\mathrm{thin}}$ and $^{(\mathrm{FT})}X_{ab}$, a choice we can justify by our experience that it works well in the numerical implementation of the moment calculation \flash~\cite{Grohs:2022fyq}. But given the possibility of other choices, we shall discuss the robustness of our results with regard to choosing different options for the closure (see Sec.~\ref{subsubsec:comparison_num}).

\section{Linear stability analysis}
\label{sec:LSA}

Now that we have defined the quantum moments, how they evolve, and the closure, we seek to determine whether a fast flavor instability can take place for a given set of classical moments, i.e., if we were to place the values for the classical moments on the flavor-diagonal parts of quantum number densities and fluxes, whether neutrinos would undergo flavor transformation. We can answer this question by performing a linear stability analysis of the system of equations~\eqref{eq:QKE_moment_N}--\eqref{eq:QKE_moment_Fbar}. To do so, we take purely flavor diagonal reference matrices $N_0= \mathrm{diag}(N_{ee},N_{xx})$ and $F_0^i= \mathrm{diag}(F^{i}_{ee},F^i_{xx})$, and perturb them with sinusoidal off-diagonal terms.  We only use off-diagonal perturbations because the instability is described by the exponential growth of the off-diagonal components until they reach the same order of magnitude as the diagonal ones, which corresponds to the saturation of the instability. Note that in an actual physical system, these perturbations are seeded by the vacuum term in the QKEs. However, given the large difference of scales between the vacuum and self-interaction parts of the Hamiltonian, it is sufficient here to take ad hoc perturbations and neglect the vacuum term altogether. We also verified in numerical simulations that the linear phase of the instability is equivalent in both types of perturbations.

The perturbed density and fluxes thus read:\footnote{In practice, perturbations must be Hermitian, which is ensured by adding the appropriate complex conjugates to the off-diagonal components, e.g., $N_{ex}=A_{ex}\exp[-\i(\Omega t-\vec{k}\cdot\vec{x})] + A^*_{xe}\exp[+\i(\Omega^* t-\vec{k}\cdot\vec{x})]$. We do not write them here for brevity, since linear stability analysis handles separately the different exponential terms.}
\begin{align}
    \label{eq:perturb_N}
    N &= \begin{pmatrix}
            N_{ee} & A_{e x}e^{- \i (\Omega t - \vec{k} \cdot \vec{x})} \\
            A_{x e}e^{- \i (\Omega t - \vec{k} \cdot \vec{x})} & N_{x x}
        \end{pmatrix} \, , \\
    \label{eq:perturb_F}
    F^j &= \begin{pmatrix}
            F^j_{ee} & B^j_{e x}e^{- \i (\Omega t - \vec{k} \cdot \vec{x})} \\
            B^j_{x e}e^{- \i (\Omega t - \vec{k} \cdot \vec{x})}& F^j_{xx}
        \end{pmatrix} \, ,
\end{align}
with similar expressions for antineutrinos. The perturbations of the pressure tensor are obtained consistently from the choice of closure, as we now show. Our goal is to obtain the relation $\Omega(\vec{k})$ and, more precisely, to look for modes where $\Im(\Omega) > 0$. To that end, we linearize the QKEs at first order in $\{A_{ex},\vec{B}_{ex},\bA_{xe},\vec{\bB}_{xe}\}$ (identical equations can be obtained with the variables $\{A_{xe},\vec{B}_{xe},\bA_{ex},\vec{\bB}_{ex}\}$, but this provides completely redundant information).

\subsection{Derivation of linear stability equations}

For convenience, we introduce some auxiliary quantities reflecting the ELN-XLN moments that appear in the self-interaction Hamiltonian:
\begin{equation}
    \begin{aligned}
        \nbDelta_N &\equiv \sqrt{2} G_F (\, \nbN_{ee} - \nbN_{xx} ) \, , \\
        \vec{\nbDelta}_F &\equiv \sqrt{2} G_F (\, \vec{\nbF}_{ee} - \vec{\nbF}_{xx} )  \, .
    \end{aligned}
\end{equation}
They allow us to define the “primed” frequency $\Omega'$ and wavevector $\vk'$:
\begin{equation}
\label{eq:def_omp_kp}
    \begin{aligned}
\Omega' &\equiv \Omega - (\Delta_N - \bDelta_N) \, , \\
\vec{k}' &\equiv \vec{k} - (\vec{\Delta}_F - \vec{\bDelta}_F) \, .
\end{aligned}
\end{equation}
These shifted quantities label a mode in the co-rotating frame in flavor space, which takes out the overall rotation of the flavor isospins due to the net ELN-XLN current (see, e.g.,~\cite{Dasgupta:2018ulw,Duan:2005cp}). 

The first equations of the moment hierarchies, Eqs.~\eqref{eq:QKE_moment_N} and \eqref{eq:QKE_moment_Nbar}, are then linearized as:
\begin{align}
    \Omega' A_{ex} \! -  \vec{k}' \!  \cdot \!  \vec{B}_{ex}  &= - \Delta_N (A_{ex} - \bA_{xe}) + \vec{\Delta}_F \! \cdot \! (\vec{B}_{ex} - \vec{\bB}_{xe})  , \label{eq:QKE_lin_A} \\
    \Omega' \bA_{xe} \! -  \vec{k}' \! \cdot \! \vec{\bB}_{xe}  &= - \bDelta_N (A_{ex} - \bA_{xe}) + \vec{\bDelta}_F \! \cdot \! (\vec{B}_{ex} - \vec{\bB}_{xe})  . \label{eq:QKE_lin_Abar}
\end{align}
To linearize Eqs.~\eqref{eq:QKE_moment_F} and \eqref{eq:QKE_moment_Fbar} we need to invoke the closure, discussed in Sec.~\ref{subsec:closure}. We write the results in the following form: 
\begin{align}
    \Omega' B_{ex}^i - \Pi^{ij}_{ex} k'_j A_{ex} &= - \Delta_F^i (A_{ex} - \bA_{xe}) \nonumber \\ 
    &\qquad \qquad + \Delta_P^{ij}(B_{ex}^j - \bB_{xe}^j) \, , \label{eq:QKE_lin_B} \\
    \Omega' \bB_{xe}^i - \bPi^{ij}_{xe} k'_j \bA_{xe} &= - \bDelta_F^i (A_{ex} - \bA_{xe}) \nonumber \\ &\qquad \qquad + \bDelta_P^{ij} (B_{ex}^j - \bB_{xe}^j) \, . \label{eq:QKE_lin_Bbar}
\end{align}
The expressions of $\Pi^{ij}$ and $\Delta_P^{ij}$ are closure-dependent. Namely,
\begin{align}
    \Delta_P^{ij} &\equiv \sqrt{2} G_F \left(P^{ij}_{ee} - P^{ij}_{xx}\right) \nonumber \\
    &= \sqrt{2} G_F \left(\Pi^{ij}_{ee} N_{ee} - \Pi^{ij}_{xx} N_{xx}\right) \, .
\end{align}
The “\emph{pressure shape}” matrix $\Pi^{ij}$ is parametrized from Eq.~\eqref{eq:Pij_closed_quantum} as:
\begin{equation}
    \Pi^{ij}_{ab} = \frac{3 X_{ab} - 1}{2} w_{ab}^{ij} + \frac{3(1 - X_{ab})}{2} \frac{\delta^{ij}}{3} \, ,
\end{equation}
such that $P^{ij}_{ab}=\Pi^{ij}_{ab}N_{ab}$. The optically thick limit ($\delta^{ij}/3$) is directly obtained from~\eqref{eq:Pij_thick}, while the matrix $w_{ab}^{ij}$ depends on the choice made for the free streaming limit $P_{ab,\mathrm{thin}}^{ij}$.

For the choice $^{(\mathrm{CW})}P_{ab,\mathrm{thin}}^{ij}$ [Eq.~\eqref{eq:Pij_thin_1}], it reads:
\begin{equation}
^{(\mathrm{CW})}w^{ij}_{aa} = \frac{F_{aa}^i F_{aa}^j}{\abs*{\vec{F}_{aa}}^2} \quad \text{and} \quad ^{(\mathrm{CW})}w^{ij}_{ex} = \frac{B_{ex}^i B_{ex}^j}{\abs*{\vec{B}_{ex}}^2} \, .
\end{equation}
Indeed, the $ex$ component of the pressure tensor is then
\begin{equation}
    ^{(\mathrm{CW})}P^{ij}_{ex, \mathrm{thin}} = A_{ex} \frac{B_{ex}^i B_{ex}^j}{\abs*{\vec{B}_{ex}}^2} e^{- \i (\Omega t - \vec{k} \cdot \vec{x})} + \cdots \, ,
\end{equation}
where the other terms are subdominant for an unstable mode ($\Im(\Omega) > 0$). This term appears directly on the left-hand side of the QKE~\eqref{eq:QKE_moment_F} and thus on the left-hand side of~\eqref{eq:QKE_lin_B}.

For the choice $^{(\mathrm{FT})}P_{ab,\mathrm{thin}}$ [Eq.~\eqref{eq:Pij_thin_2}], it reads:
\begin{equation}
\label{eq:wij_2}
^{(\mathrm{FT})}w^{ij}_{ab} = \frac{(F_{ee}^i + F_{xx}^i) (F_{ee}^j + F_{xx}^j)}{\abs*{\vec{F}_{ee}+\vec{F}_{xx}}^2} \ \ \ \forall \{a,b\} \, .
\end{equation}
All expressions are similar for antineutrinos. In the first case, there is an ambiguity regarding the definition of $^{(\mathrm{CW})}w_{ex}^{ij}$, since it makes equation~\eqref{eq:QKE_lin_B} non-linear in the perturbations. We thus set the value of the off-diagonal component $w_{ex}^{ij}$ to \eqref{eq:wij_2} whether the diagonals are assumed to be set by the (FT) or (CW) conditions. 

\subsection{Stability matrix}

We compile the set of equations~\eqref{eq:QKE_lin_A}, \eqref{eq:QKE_lin_Abar}, \eqref{eq:QKE_lin_B} and \eqref{eq:QKE_lin_Bbar} into a matrix form:
\begin{equation}
\label{eq:eq_Sk_LSA}
\left(S_{\vk'} + \Omega' \mathbb{I}\right) \cdot Q = 0 \, ,
\end{equation}
where $Q = \left(A_{e x}, B_{e x}^\x, B_{e x}^\y, B_{e x}^\z, \bA_{xe}, \bB_{xe}^\x, \bB_{xe}^\y, \bB_{xe}^\z\right)^T$ is the vector of perturbation amplitudes, and $\mathbb{I}$ is the $8 \times 8$ identity matrix. For completeness, we give the full expression of $S_{\vk'}$, that we call the “stability matrix”, separated between a $\vk'$-independent part and a $\vk'$-dependent one: 
\begin{equation}
    \label{eq:stability_matrix}
    S_{\vk'} = \begin{pmatrix} \widetilde{\Delta} & - \widetilde{\Delta} \\ \widetilde{\overline{\Delta}} & - \widetilde{\overline{\Delta}} \end{pmatrix} - \Sigma_{\vk'} \, ,
\end{equation}
where
\begin{equation}
    \widetilde{\Delta} \equiv \left(\begin{array}{c|c}
 \Delta_N & \quad - \vec{\Delta}_F^T \quad  \\ \hline & \\
\vec{\Delta}_F &  - \left[{\Delta_P}\right]  \\ & \\
 \end{array} \right) \, , 
\end{equation}
with an analogous expression for $\widetilde{\overline{\Delta}}$, and

\begin{equation}
 \Sigma_{\vk'} \equiv \left( 
 \begin{array}{c|c||c|c} 
0 & \quad \vec{k}'^T \quad & 0 & \quad \vec{0}^T \quad  \\ \hline & & & \\
\left[\Pi_{ex}\right] \cdot \vec{k}' & [0] & \vec{0} & [0] \\ & & & \\ \hline \hline
0 & \vec{0}^T & 0 & \vec{k}'^T \\ \hline & & & \\
\vec{0} & [0] & \left[\overline{\Pi}_{xe}\right]  \cdot \vec{k}' & [0] \\ & & & 
\end{array}\right) \, .
\end{equation}
For clarity, we write $3 \times 3$ matrices with brackets ($[\cdots]$) and $4\times4$ matrices with a tilde ($\widetilde{\cdots}$).

Non-zero solutions of the system of equations~\eqref{eq:eq_Sk_LSA} are obtained by numerically solving:
\begin{equation}
\det \left(S_{\vk'} + \Omega' \mathbb{I} \right) = 0 \, .
\end{equation}
In other words, the solutions $\Omega'(\vk')$ are the negative of the eigenvalues of $S_{\vk'}$. Should $\Im(\Omega') > 0$, and equivalently $\Im(\Omega) > 0$ (since the constant shift between $\Omega$ and $\Omega'$, see Eq.~\eqref{eq:def_omp_kp}, is purely real), the associated mode would be unstable and exponentially growing at the rate $\Im(\Omega)$. For a given $\vec{k}$ — and correspondingly $\vec{k}'$ given by Eq.~\eqref{eq:def_omp_kp} — there are eight eigenvalues of $S_{\vk'}$, of which the eigenmode with the largest value of $\Im(\Omega)$ will dominate. We scan for all values of $\vec{k}$, such that the overall fastest growing mode is:
\begin{equation}
    \Im(\Omega)_\mathrm{max} \equiv \underset{\vec{k}}{\mathrm{max}} \big\{ \, \Im\left[\Omega(\vec{k})\right] \big\} \, ,
\end{equation}
corresponding to a wavevector $\vec{k}_\mathrm{max}$.

\paragraph*{Consistency check —} When using the LSA described above we find that in some cases, “spurious” solutions appear, generally at large values of $k$. Such modes are not related to a problem of angular discretization (see~\cite{Morinaga:2018aug}) due to the inherent angular-integrated nature of the moment method. However, it appears that these modes generally correspond to eigenvectors that show strong directional structure beyond what can be adequately described by only two moments. We present in Appendix~\ref{app:multi_angle_LSA} a simplified multi-angle LSA which allows us to estimate the angular distribution of the unstable mode eigenvectors based on the moment LSA results. If this angular distribution is not smooth enough, a truncated two-moment method with a (necessarily) imperfect closure is not expected to accurately describe the situation. We thus conservatively discard such modes based on this multi-angle LSA “consistency check” by applying a cutoff on the derivative of the multi-angle eigenvector angular distribution [see Eq.~\eqref{eq:cutoff}]. This procedure is supported by our results: for the physical conditions studied in~\cite{Grohs:2022fyq} and discussed in Sec.~\ref{subsubsec:comparison_num}, we find this type of spurious modes at large $k$, but the fastest growing mode we obtain by eliminating those modes agrees with numerical results. More generally, no such spurious modes are observed in the fully non-linear calculations, whether they are carried out with moments (\flash) or particle-in-cell multi-angle (\emu) methods. Nevertheless, some unstable modes might be missed by the LSA given the approximate nature of this test (see also the discussion on ELN crossings in Sec.~\ref{subsec:ELN_crossing}).

\section{Test case}
\label{sec:fiducial}

In order to illustrate how we use the moment linear stability and how it compares to actual calculations, we consider in this section a simple test case with the parameters quoted in Table~\ref{tab:fiducial}. It physically represents two extended beams of electron neutrinos and antineutrinos on average propagating in opposite directions $\pm \z$, and was dubbed in~\cite{Richers_emu,Grohs:2023pgq} the “Fiducial” test case.

\begin{table}[!ht]
    \begin{tabular}{c|c}
    Moment & Value \\
    \hline 
        $N_{ee}$ & $4.89 \times 10^{32} \ \mathrm{cm}^{-3}$ \\
        $\bN_{ee}$ & $4.89 \times 10^{32} \ \mathrm{cm}^{-3}$ \\
        $N_{xx} = \bN_{xx}$ & $0$ \\ \hline
        $\vec{F}_{ee}/N_{ee}$ & $(0,0, \ \, 1/3 \ )$ \\
        $\vec{\bF}_{ee}/\bN_{ee}$ & $(0,0,-1/3)$ \\
        $\vec{F}_{xx} = \vec{\bF}_{xx}$ & $(0,0, \ \ \ 0 \ \ \, )$ \\ 
    \end{tabular}
    \caption{\label{tab:fiducial} Parameters for the Fiducial test case. The electron (anti)neutrino densities are of the typical order of magnitude of neutron star merger values, which ensures the appropriate scaling of the fast flavor instability growth rate.}
\end{table}

\begin{figure}[!ht]
    \centering      
    \includegraphics{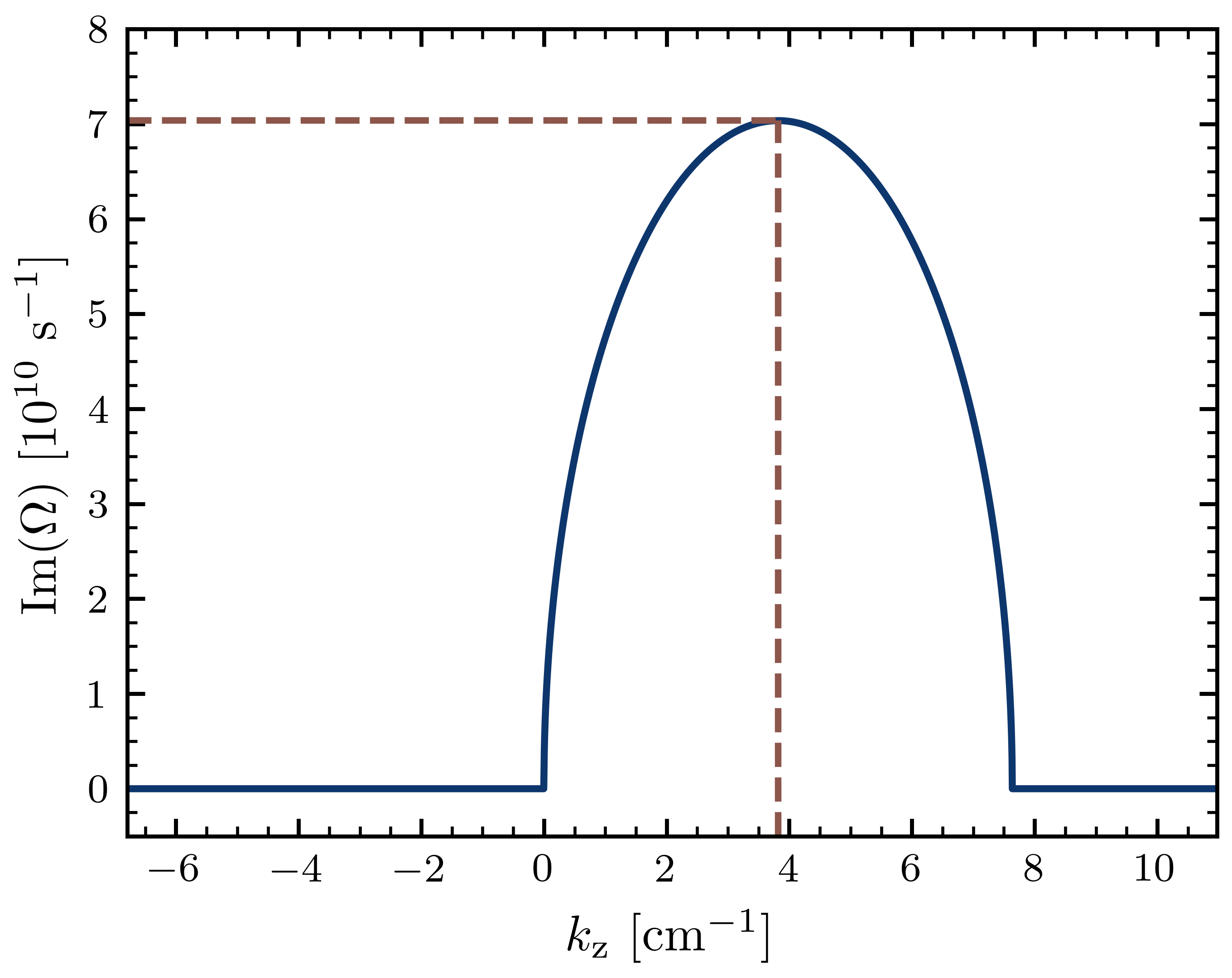}
    \caption{\label{fig:LSA_fiducial} Instability growth rate for different wavenumbers in the Fiducial test case. The coordinates of the overall maximum are given in Eq.~\eqref{eq:max_fiducial}.}
\end{figure}

\paragraph*{LSA prediction ---} The wavevector of the fastest growing mode is parallel to $\vec{z}$, which is not surprising given the azimuthal symmetry around $\vec{z}$ in this example. We show in Fig.~\ref{fig:LSA_fiducial} the maximum growth rate for each value of $\vec{k} = (0,0,k_\z)$, and we identify the overall fastest growing mode:
\begin{equation}
\label{eq:max_fiducial}
k_\mathrm{max} \simeq 3.82 \, \mathrm{cm^{-1}} \quad \text{;} \quad \Im(\Omega)_\mathrm{max} \simeq 7.04 \times 10^{10} \, \mathrm{s^{-1}} \, .
\end{equation}
This test case being one-dimensional and satisfying many symmetries (see values in Table~\ref{tab:fiducial}), it can be solved analytically.\footnote{Note that the fastest growing mode characteristics for the Fiducial test in \cite{Richers_emu} were calculated assuming a distribution linear in the cosine of the angle from the $\z$ axis instead of a maximum entropy distribution, so while the results here are similar, they do not match exactly.} With the notations of Sec.~\ref{sec:LSA}, we get
\[
\frac{k'_\mathrm{max}}{\sqrt{2} G_F N_{ee}} = \frac{28}{51}   \quad \text{and} \quad
\frac{\Im(\Omega)_\mathrm{max}}{\sqrt{2} G_F N_{ee}} = \frac{62}{9 \sqrt{85}} \, ,
\]
in perfect numerical agreement with Eq.~\eqref{eq:max_fiducial}.

\medskip

\paragraph*{Comparison with QKE simulations ---} We compare this LSA prediction with two simulations: a particle-in-cell multi-angle calculation, \emu~\cite{Richers_emu}, and a moment calculation, \flash~\cite{Grohs:2022fyq,Grohs:2023pgq}. \emu simulates a large number of computational particles moving in many individual directions. Each of these particles is described by a position, momentum, and a flavor density matrix, which are evolved by solving directly Eqs.~\eqref{eq:QKE_rho}--\eqref{eq:QKE_rhobar} using a particle-in-cell method with second-order shape functions and a global fourth-order time integration scheme. The initial flavor diagonal components of the density matrices are attributed to individual particles such that the angular moments $N_{aa}$, $\bN_{aa}$, $\vec{F}_{aa}$ and $\vec{\bF}_{aa}$ are reproduced, with the angular distribution consistent with the classical closure~\eqref{eq:chi_MEC} for $\nbnuin_e$ and $\nbnuin_x$ [see Eq.~\eqref{eq:dist_ang_ME}]. \flash is a moment code, which solves Eqs.~\eqref{eq:QKE_moment_N}--\eqref{eq:QKE_moment_Fbar} using the closure relation outlined in Sec.~\ref{subsec:closure}, choosing specifically the options~\eqref{eq:Pij_thin_1} and \eqref{eq:chi_ab}. However, \flash evolves separately the real and imaginary parts of $N_{ex}$ and $\vec{F}_{ex}$, considered as two neutrino “species”, for which Eq.~\eqref{eq:Pij_thin_1} is separately evaluated. This leads to inevitable differences with the LSA prediction. The flavor diagonal components of $N$, $\bN$, $\vec{F}$ and $\vec{\bF}$ are initially provided, similarly to the LSA. An initial random perturbation to the flavor off-diagonal components of the quantities respectively evolved in \emu (particle density matrices) and \flash (moments) seeds the fast flavor instability. For the simulations presented in this work, the domain size is $L = 8 \, \mathrm{cm}$ with $128$ grid points in each space dimension, and the \emu simulations use 378 particles per cell. Additional details on the implementation can be found in~\cite{Grohs:2023pgq,Richers_emu}.

\begin{figure*}[!ht]
    \centering \includegraphics{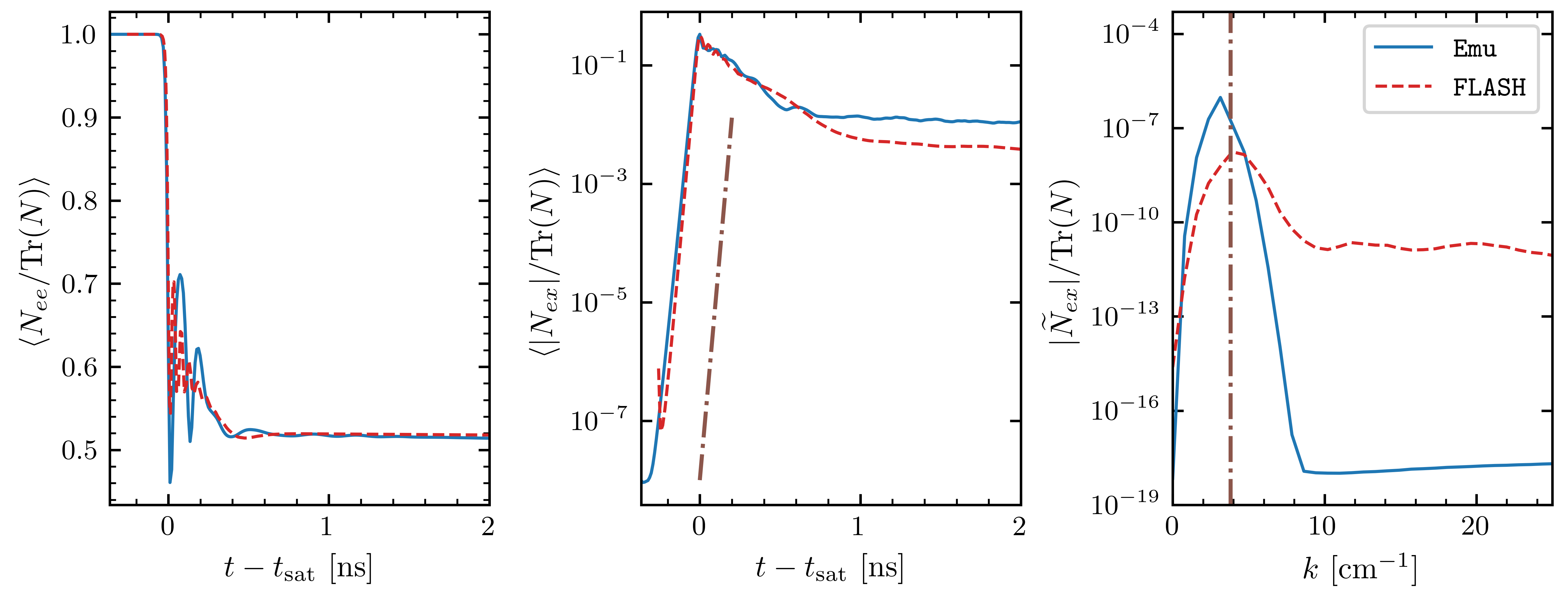}
    \caption{\label{fig:simu_fiducial} Numerical solution of three-dimensional neutrino flavor transformation in the Fiducial test case, for particle-in-cell (\emu) and moment (\flash) codes, adapted from~\cite{Grohs:2023pgq}. For comparison purposes, the time origin on the left and middle plots is taken at the saturation time $t_\mathrm{sat}$ (for which $N_{ex}$ is maximal), which occurs at  different times in the \emu and \flash simulations. \emph{Left:} evolution of the normalized electron neutrino density averaged over the simulation box, which shows fast flavor conversion. \emph{Middle:} evolution of the domain-averaged off-diagonal component of $N$, showing the exponential growth of the unstable mode. The brown line shows an exponential growth at the rate given by the linear stability analysis prediction~\eqref{eq:max_fiducial}. \emph{Right:} magnitude of the spatial Fourier transform of the off-diagonal component ($\widetilde{N}_{ex}$) against wavenumber $k$, taken $0.1 \, \mathrm{ns}$ before saturation. The dash-dotted line is the linear stability analysis prediction for $k_\mathrm{max}$.}
\end{figure*}

Results are shown in Fig.~\ref{fig:simu_fiducial}. We find a very good agreement between the prediction and both numerical results~\cite{Grohs:2023pgq}:
\begin{equation}
\label{eq:max_fiducial_simu}
\begin{aligned}
k_\mathrm{max}^\flash &\simeq 3.9(4) \, \mathrm{cm^{-1}} \ \ \text{;}  & \Im(\Omega)_\mathrm{max}^\flash \simeq&& \! \! \! 7.1 \times 10^{10} \, \mathrm{s^{-1}} \, , \\
k_\mathrm{max}^\emu &\simeq 3.1(4) \, \mathrm{cm^{-1}} \ \ \text{;}  & \Im(\Omega)_\mathrm{max}^\emu \simeq&& \! \! \! 6.3 \times 10^{10} \, \mathrm{s^{-1}} \, .
\end{aligned}
\end{equation}
It should be noted that although the initial conditions for \emu follow the angular distribution assumed by the Minerbo closure, the evolution thereafter is independent of that assumption. It is, therefore, reasonable that the results differ slightly from the analytic stability analysis that requires the Minerbo closure be followed at all times. As mentioned before, \flash does assume a Minerbo closure throughout, but its implementation is slightly different from the assumptions made in the linear stability analysis (see \cite{Grohs:2023pgq} for details). An improved treatment of the closure in \flash would be required for us to study strict convergence to the analytic result.

The initial conditions involve neutrino densities that are typical of a NSM-like environment. Both \emu and \flash simulations, in agreement with the LSA prediction, show that the growing mode has a characteristic wavelength on the scale of millimeters (right panel of Fig.~\ref{fig:simu_fiducial}) and a growth phase that lasts for about $\sim 0.1 \, \mathrm{ns}$ (left two panels of Fig.~\ref{fig:simu_fiducial}).

We also find good agreement with the “90-Degree” and “Two Thirds” test cases considered in~\cite{Grohs:2023pgq}. The growth rate predicted by LSA is $15 \, \%$ larger than the \emu value and $7 \, \%$ smaller than the \flash value in the 90-Degree case, and is $30 \, \%$ larger than the \emu value and $20 \, \%$ smaller than the \flash value in the Two Thirds case.


\section{Fast flavor instabilities across a post-merger remnant}
\label{sec:NSM}

We now turn to the application of moment linear stability analysis in a neutron star merger environment. Full hydrodynamics simulations including an accurate treatment of neutrino flavor oscillations are out of reach for now, but a moment treatment of neutrino transport has an interesting potential in terms of reducing the computation time at the cost of an approximate closure relation. We thus want to assess whether moment methods would accurately describe the phenomenon of fast flavor conversion, a possibility evidenced in numerical simulations by Refs.~\cite{Grohs:2022fyq,Grohs:2023pgq}. To this end, we perform a linear stability analysis on the results of the classical general relativistic two-moment radiation hydrodynamics simulation of the merger of two $1.2 \, M_\odot$ neutron stars from Refs.~\cite{Foucart:2015gaa,Foucart:2016rxm}. This simulation provides the first two moments of the neutrino distributions throughout the NSM remnant, assuming equal numbers of heavy lepton flavor neutrinos and antineutrinos. In other words, our starting point is the set of classical moments $\{N_{ee},\bN_{ee},N_{xx}=\bN_{xx},\vec{F}_{ee},\vec{\bF}_{ee},\vec{F}_{xx}=\vec{\bF}_{xx}\}$ for each point on a three-dimensional grid ($201 \times 201 \times 101$ points) corresponding to physical dimensions ($136 \, \mathrm{km} \times 136 \, \mathrm{km} \times 68 \, \mathrm{km}$) from a snapshot taken $5 \, \mathrm{ms}$ after the merger.

Note that we distinguish between global spatial coordinates in the NSM simulation $(X,Y,Z)$, and local coordinates which are defined differently at each point $(\x, \y, \z)$. The local coordinates are chosen such that the lepton number flux vector $\vec{F}_{ee} - \vec{\bF}_{ee}$ is aligned with $\vec{z}$.\footnote{As a consequence, the local $(\x, \y, \z)$ coordinates in different points of the NSM simulation are \emph{not} related.}
In addition, $x$ denotes the heavy lepton flavor.

\subsection{Results from linear stability analysis}
\label{subsec:NSM_results}

\begin{figure}[!ht]
    \centering  
    \includegraphics[width=\columnwidth]{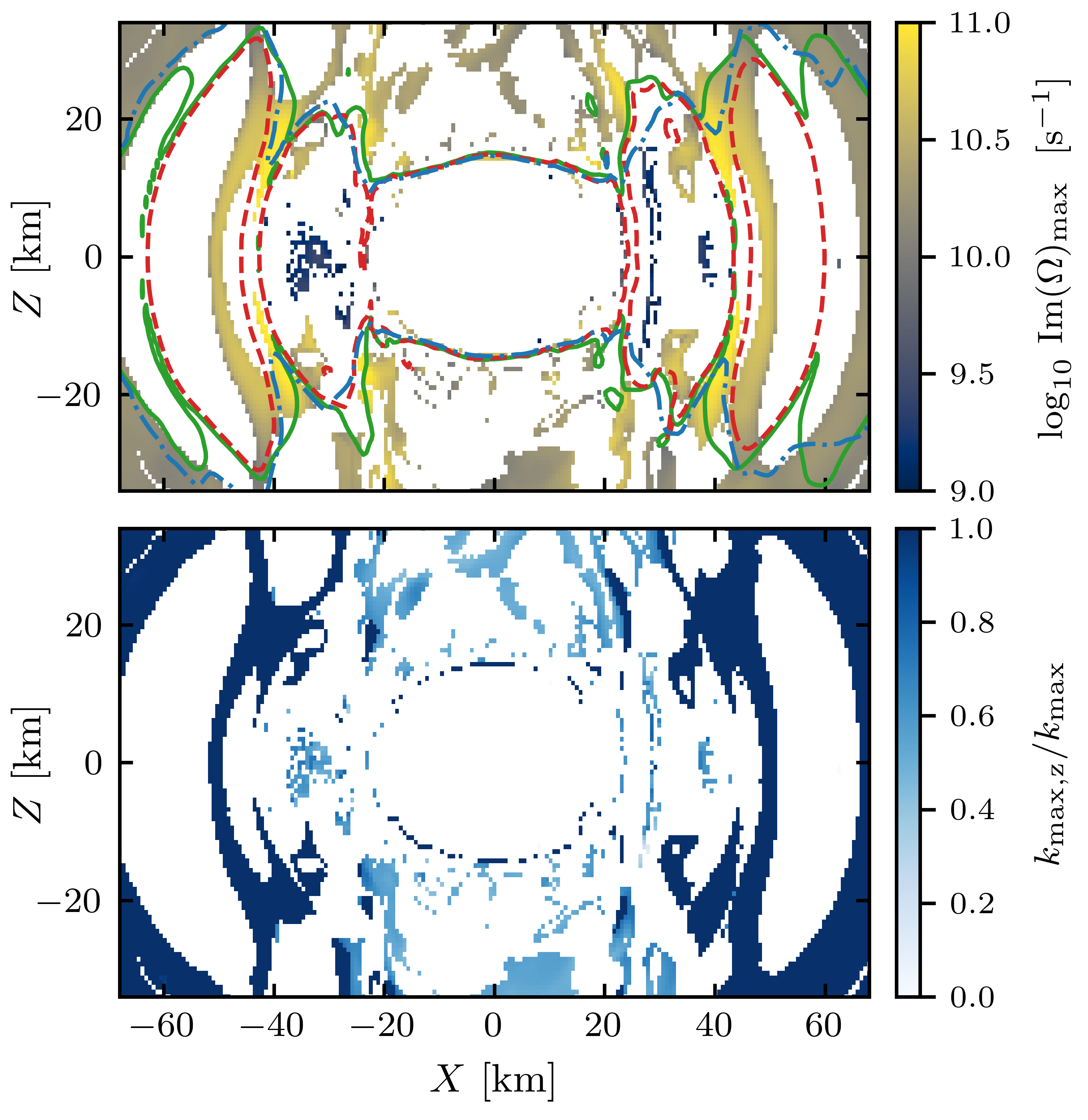}
    \caption{\label{fig:slice_3D} Predicted characteristics of the FFI fastest growing mode on the slice $\{ Y = 0 \, \mathrm{km} \}$ of the NSM simulation, obtained from moment LSA. \emph{Top:} instability growth rate. Contours of equal flux factor $\hat{f} = 0.2$ are represented for $\nu_e$ (solid green line), $\bar{\nu}_e$ (dashed red line) and $\nu_x$ (dash-dotted blue line). \emph{Bottom:} normalized vertical component of $\vec{k}_\mathrm{max}$, showing the predominance of modes aligned with the ELN flux direction. Regions with $\vec{k}_{\mathrm{max}} \nparallel \vec{z}$ are not expected to be unstable (see Sec.~\ref{subsec:ELN_crossing}), and are thus expected to be false LSA positives.}
\end{figure}

At a given location in a NSM, we can generally search for instabilities associated with any wavevector $\vec{k}$ and find the overall fastest growing mode. We show the result for a slice taken across the disk ($Y = 0 \, \mathrm{km}$) in Fig.~\ref{fig:slice_3D}. The growth rate (typically a few $10^{10} \, \mathrm{s}^{-1}$) is shown in the top panel, and we assess the direction of $\vec{k}_\mathrm{max}$ in the bottom panel. Given our choice of local coordinates, $k_{\mathrm{max},\z}$ is the component along the ELN flux direction.

In addition, we show on the top panel the location of the “neutrinospheres” for each species of neutrinos ($\nu_e$, $\bar{\nu}_e$ and $\nu_x$). More precisely, as the optical depth is not available in the simulation data (and not well defined in the complex 3D geometry of a merger remnant), we use a proxy for the transition between trapped and optically-free regions by considering the flux factor threshold $\hat{f} = 0.2$, which roughly corresponds to an optical depth $\tau \sim 2/3$ \cite{Foucart:2016rxm}. These contours reveal the very aspherical structure of the decoupling regions, corresponding to the neutron star core and the overdense tidal arms. The complex structure within the accretion torus and the relativistic orbital velocities justify performing our analysis of the detailed neutrino distributions produced by multidimensional relativistic radiation transport. The orbital velocities, in particular, cause radiation emitted from the accretion torus to be beamed preferentially along the surface of the torus, causing the radiation in the polar regions to differ significantly from models of neutrino emission for stationary fluid.

The white regions in Fig.~\ref{fig:slice_3D} show locations where the FFI is not present. Among the unstable locations, we can distinguish two regions: in most of the slice, $\vec{k}_\mathrm{max} \parallel \vec{z}$ (dark blue in the bottom panel), while notably in the polar regions the transverse component seems to be more important (light blue).

As a matter of fact, the modes which are not aligned with $\vec{z}$ (see bottom panel of Fig.~\ref{fig:slice_3D}) are particularly dependent on the consistency check introduced at the end of Sec.~\ref{sec:LSA}, and furthermore they correspond to regions where we do not expect an FFI to occur based on the ELN crossing test (see Sec.~\ref{subsec:ELN_crossing}). For these reasons, we expect our method to provide positive results when modes are found along the ELN flux direction, to which we will restrict in the remainder of this paper.

\begin{figure}[!ht]
    \centering  
    \includegraphics[width=\columnwidth]{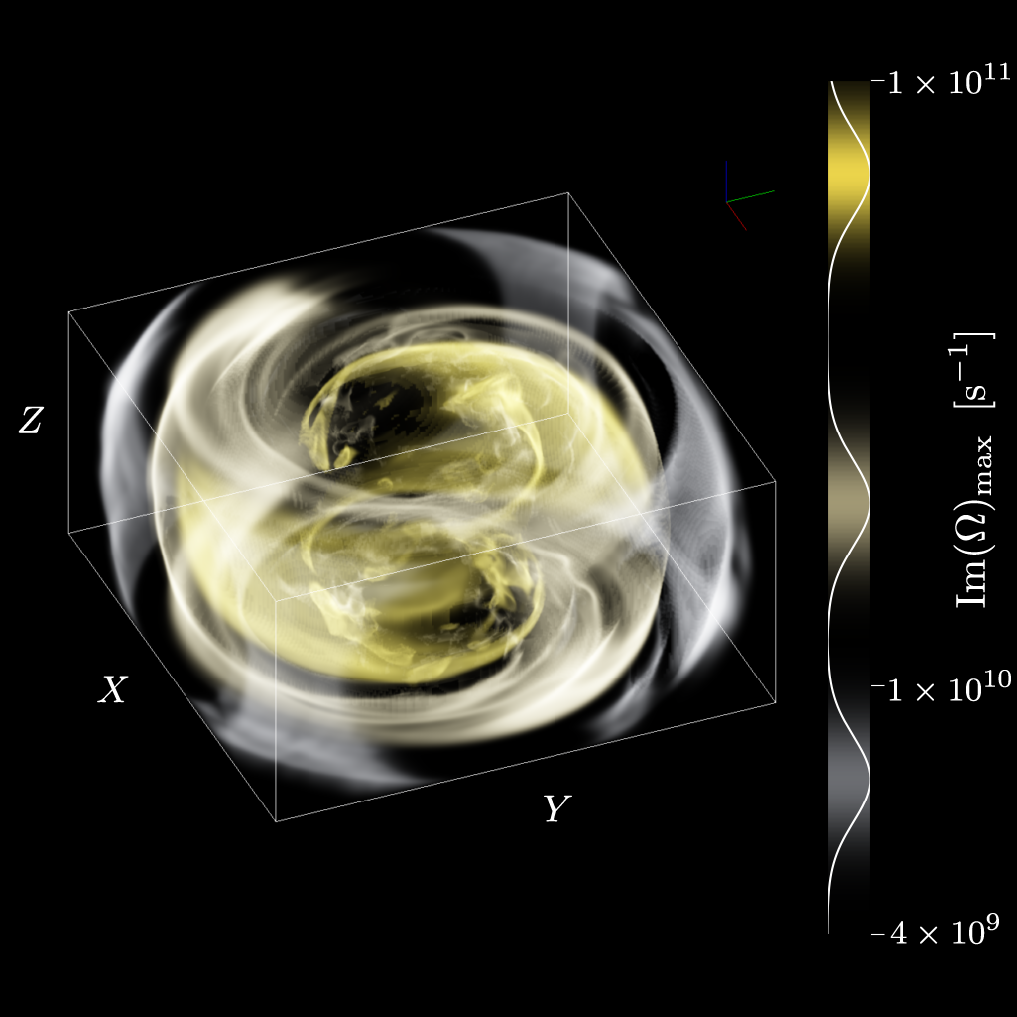}
    \caption{\label{fig:results_3D} Volume rendering of the fast flavor instability growth rate in the NSM simulation from~\cite{Foucart:2015gaa,Foucart:2016rxm}, taken $5 \, \mathrm{ms}$ post merger. Three colored contours are centered respectively around the growth rate values $7 \times 10^9 \, \mathrm{s^{-1}}$, $2 \times 10^{10} \, \mathrm{s^{-1}}$, and $7 \times 10^{10} \, \mathrm{s^{-1}}$.}
\end{figure}

When applying this moment LSA more generally to every point of the NSM simulation, we find — in agreement with previous studies — that fast flavor instabilities are ubiquitous in such environments~\cite{WuTamborra}; as depicted in the 3D volume rendering of the instability growth rate on Fig.~\ref{fig:results_3D}.

\begin{figure*}[!ht]
    \centering    \includegraphics{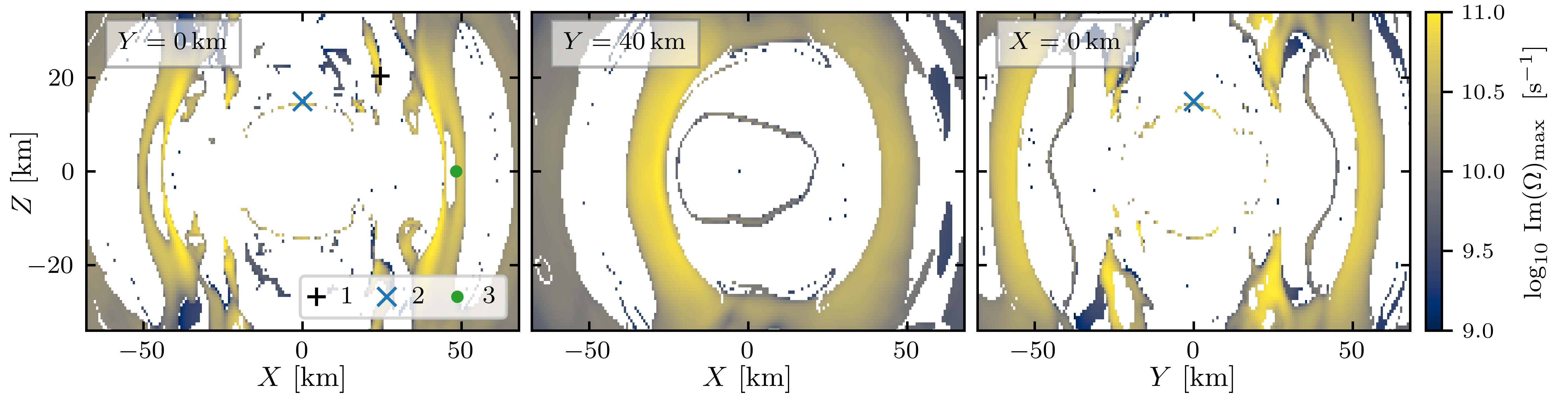}
    \caption{\label{fig:slices_NSM} Instability growth rate from moment linear stability analysis, for various slices of the NSM simulation. \emph{Left:} slice across the equatorial plane at $Y = 0 \, \mathrm{km}$. \emph{Middle:} off-center slice across the equatorial plane, at $Y \simeq 40 \, \mathrm{km}$. \emph{Right:} centered slice orthogonal to the first one, at $X = 0 \, \mathrm{km}$. We identify the three points studied in~\cite{Grohs:2023pgq} and discussed in Sec.~\ref{subsubsec:comparison_num}.}
\end{figure*}

\subsubsection{Instability timescale}

\begin{figure}[!ht]
    \centering  \includegraphics{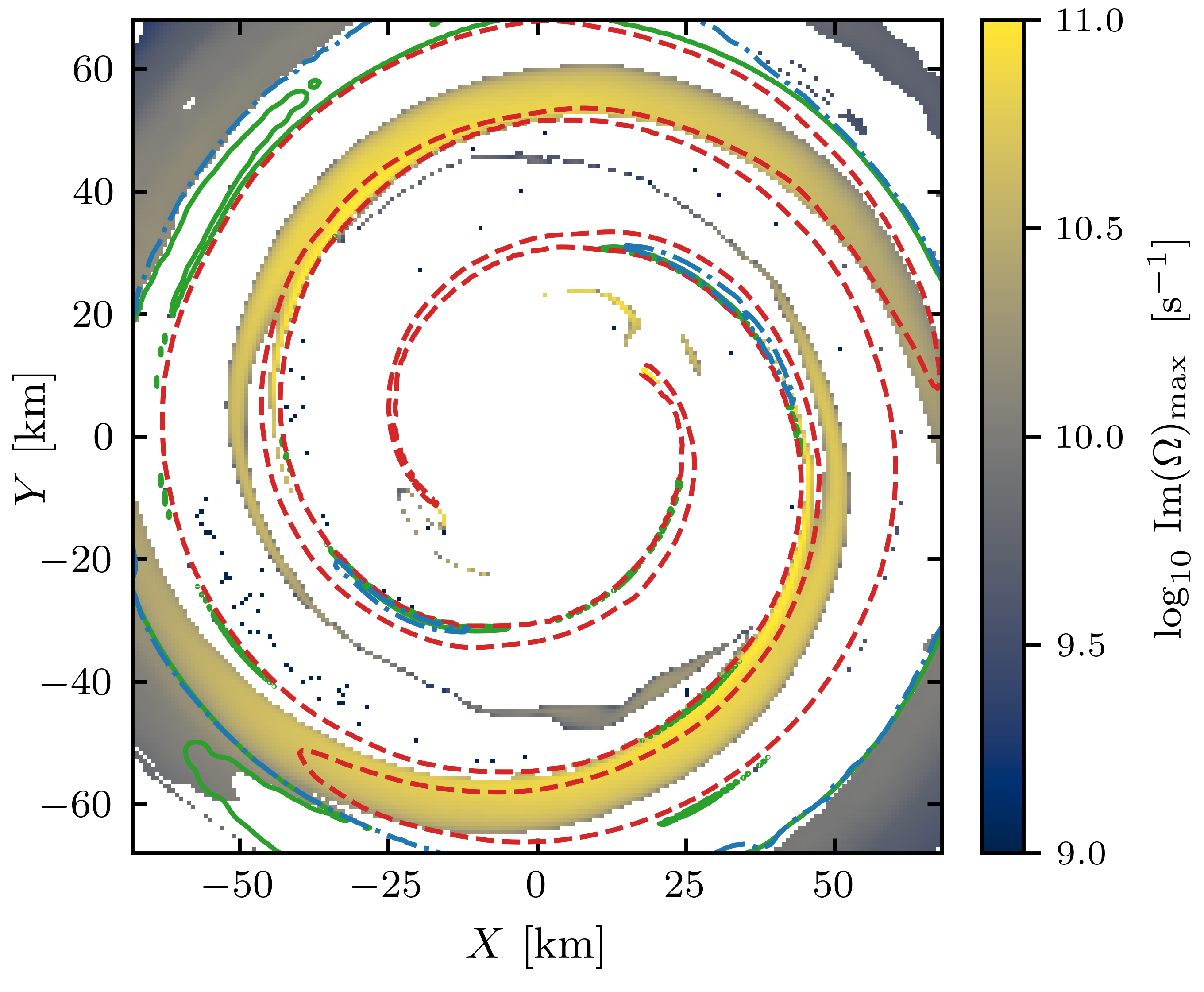}    \caption{\label{fig:Growth_equatorial} Instability growth rate obtained from moment linear stability analysis in the equatorial plane of the NSM simulation snapshot ($Z = 0 \, \mathrm{km})$. Regions of FFI follow the spiral density wave structure of the nascent accretion disk, as indicated by the $\hat{f} = 0.2$ flux factor contours, which are represented for $\nu_e$ (solid green line), $\bar{\nu}_e$ (dashed red line) and $\nu_x$ (dash-dotted blue line).}
\end{figure}

We can identify on Fig.~\ref{fig:results_3D} different unstable regions with generally higher growth rates as we get closer to the neutron star merger remnant. To visualize these results more quantitatively, we plot on Fig.~\ref{fig:slices_NSM} the quantity $\mathrm{Im}(\Omega)_\mathrm{max}$ for different vertical slices (see caption) and on Fig.~\ref{fig:Growth_equatorial} for an equatorial slice ($Z = 0 \, \mathrm{km}$). Typical values of the growth rate range from $10^{10} \, \mathrm{s^{-1}}$ to $10^{11} \, \mathrm{s}^{-1}$, corresponding to very short timescales $\sim 0.01 - 0.1 \, \mathrm{ns}$. 

As can be seen on Figs.~\ref{fig:results_3D}--\ref{fig:Growth_equatorial}, we observe a characteristic structure of FFI in a neutron star merger: one can identify a correlation between the unstable regions and the tidal arms of the post-merger remnant. To illustrate this fact, we represent on Fig.~\ref{fig:Growth_equatorial} the neutrinosphere proxies (surfaces of $\hat{f} = 0.2$ flux factor) in the same fashion as Fig.~\ref{fig:slice_3D} (top panel). These surfaces surround overdense tidal arms in the post-merger accretion disk. In addition, a small region of instability is found on the polar surface of the remnant, a region of particular interest since a majority of neutrinos are emitted from this surface. Nevertheless, a detailed study of this region would require us to take into account neutrino collisions, as shown for instance by the possibility of collisional instabilities evidenced in Appendix~\ref{app:collisional_inst}.

\subsubsection{Unstable mode wavelength}

In the following, we will only show results for the same slice as the left panel of Fig.~\ref{fig:slices_NSM} ($Y = 0 \, \mathrm{km}$), but the various conclusions remain general. 

\begin{figure}[!ht]
    \centering  \includegraphics{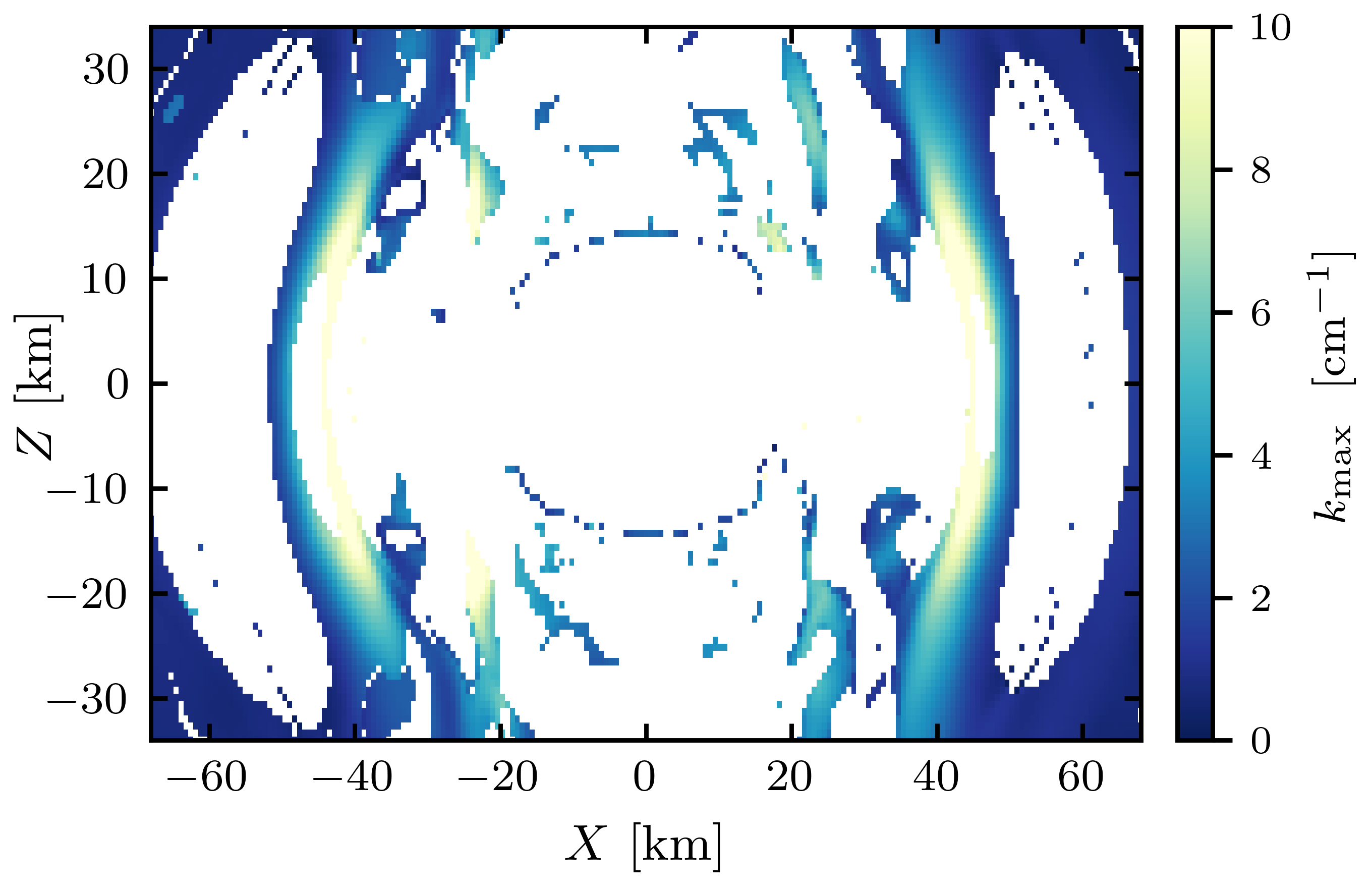}
    \caption{\label{fig:kmax_slice} Wavenumber corresponding to the fastest growing mode, across the same polar slice as the left panel of Fig.~\ref{fig:slices_NSM}. These values show that the typical lengthscale associated with FFI in a classical neutron star merger simulation is on the scale of centimeters.}
\end{figure}

The typical inverse length scale associated with the most unstable mode is given by $k_\mathrm{max}$, which is shown on Fig.~\ref{fig:kmax_slice}. Throughout the simulation, $k_\mathrm{max}$ is a few $\mathrm{cm}^{-1}$, which highlights the very small spatial scale associated to FFI compared to a typical neutron star merger simulation resolution. A comparison of the maps of Fig.~\ref{fig:kmax_slice} and Fig.~\ref{fig:slices_NSM} (left panel) shows a positive correlation between $k_\mathrm{max}$ and $\Im(\Omega)_\mathrm{max}$. That is, the larger the wavenumber, the larger the growth rate.

\medskip

\begin{figure}[!ht]
    \centering  \includegraphics{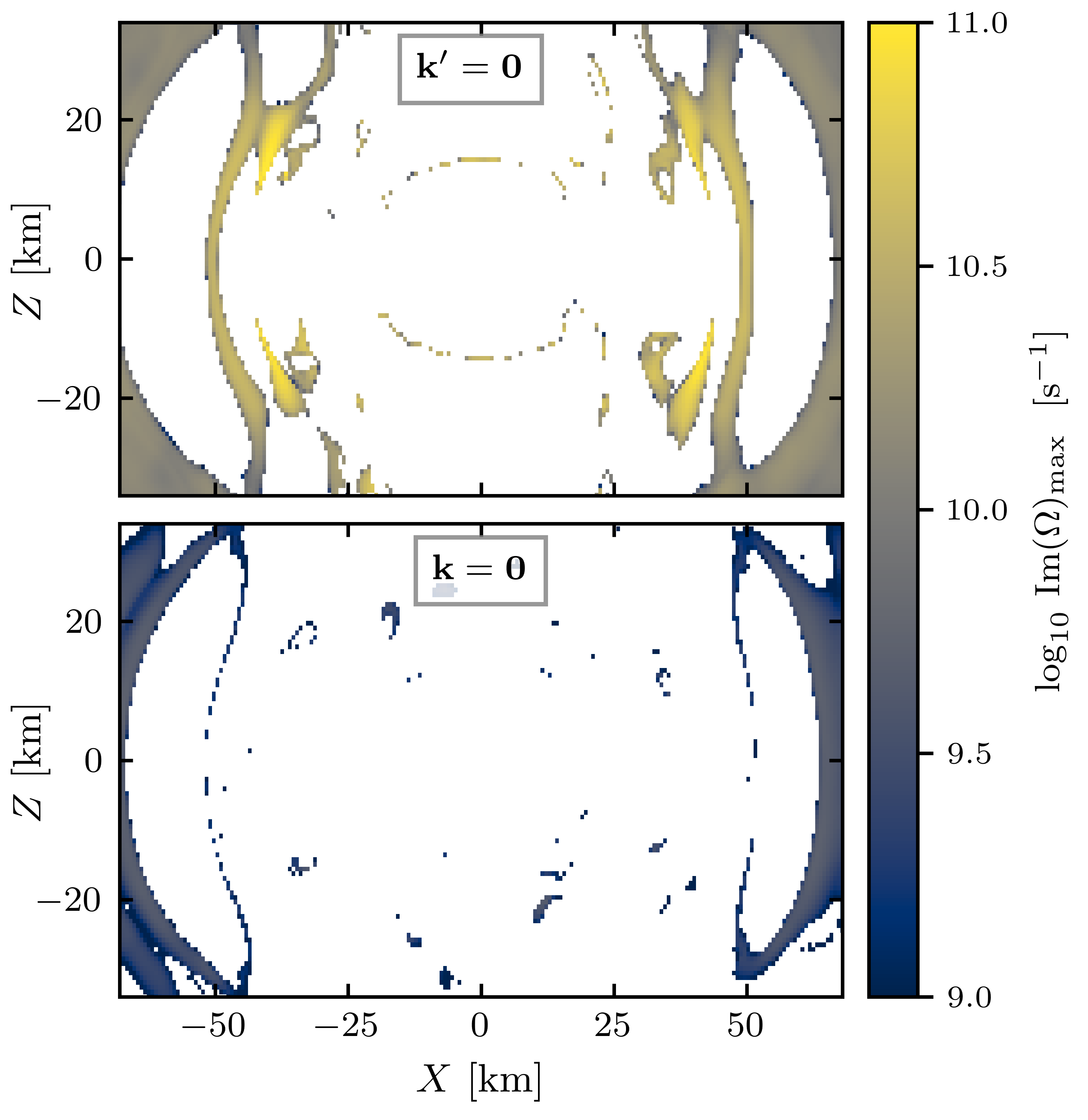}
    \caption{\label{fig:zero_hom_mode} Fast flavor instability growth rate for the NSM slice $\{Y = 0 \, \mathrm{km}\}$, obtained from moment LSA  restricted to the “zero mode” $\vec{k} = \sqrt{2} G_F (\vec{F}_{ee} - \vec{F}_{xx} - \vec{\bF}_{ee} + \vec{\bF}_{xx})$ (\emph{Top}), or to the homogeneous mode $\vec{k} = \vec{0}$ (\emph{Bottom}), to be compared with the left panel of  Fig.~\ref{fig:slices_NSM}.}
\end{figure}

\paragraph*{Zero mode ---} Some earlier studies focused solely on the so-called “zero mode”~\cite{Dasgupta:2018ulw,Glas:2019ijo,Li:2021vqj}, defined by $\vec{k}' = \vec{k} - \sqrt{2} G_F (\vec{F}_{ee} - \vec{F}_{xx} - \vec{\bF}_{ee} + \vec{\bF}_{xx}) = \vec{0}$ [see also Eq.~\eqref{eq:def_omp_kp}]. It was shown in~\cite{Richers:2022dqa} that restricting the analysis to this mode (called there the “$k_0$ test”) captures some of the unstable regions of large-scale simulations. With our analysis method, we can verify how representative this mode is in our simulation: the growth rate obtained from restricting the moment LSA to $\vec{k}' = \vec{0}$ is shown in Fig.~\ref{fig:zero_hom_mode} (top panel). This indeed confirms that some instability regions are missed, although the bulk of the unstable regions is found with a growth rate that is smaller than, but comparable to, the value when $\vec{k}' = \vec{k}'_\mathrm{max} \neq \vec{0}$.

\paragraph*{Homogeneous mode ---} Another particular mode of frequent interest is the homogeneous fast flavor instability, which corresponds to $\vec{k} = \vec{0}$. Homogeneous instabilities, like bipolar oscillations, were notably studied as part of “slow” collective oscillations (see, e.g., \cite{Duan_review,Banerjee:2011fj}), but are not expected to be particularly representative in the FFI case. Indeed, we also show in Fig.~\ref{fig:zero_hom_mode} (bottom panel) the results from linear stability analysis, restricted to $\vec{k} = \vec{0}$. Many unstable regions are missed, and in the few regions where an instability is still encountered, the growth rate is significantly smaller than the overall one, once again showing that even in those regions, a mode with $\vec{k} \neq \vec{0}$ would actually be the dominant one. It should nevertheless be noted that the domain considered remains quite close to the post-merger remnant, and including the vacuum term could notably change the results in outer regions, in particular triggering instabilities at $\vec{k} = \vec{0}$.

\begin{figure*}[!ht]
    \centering   \includegraphics{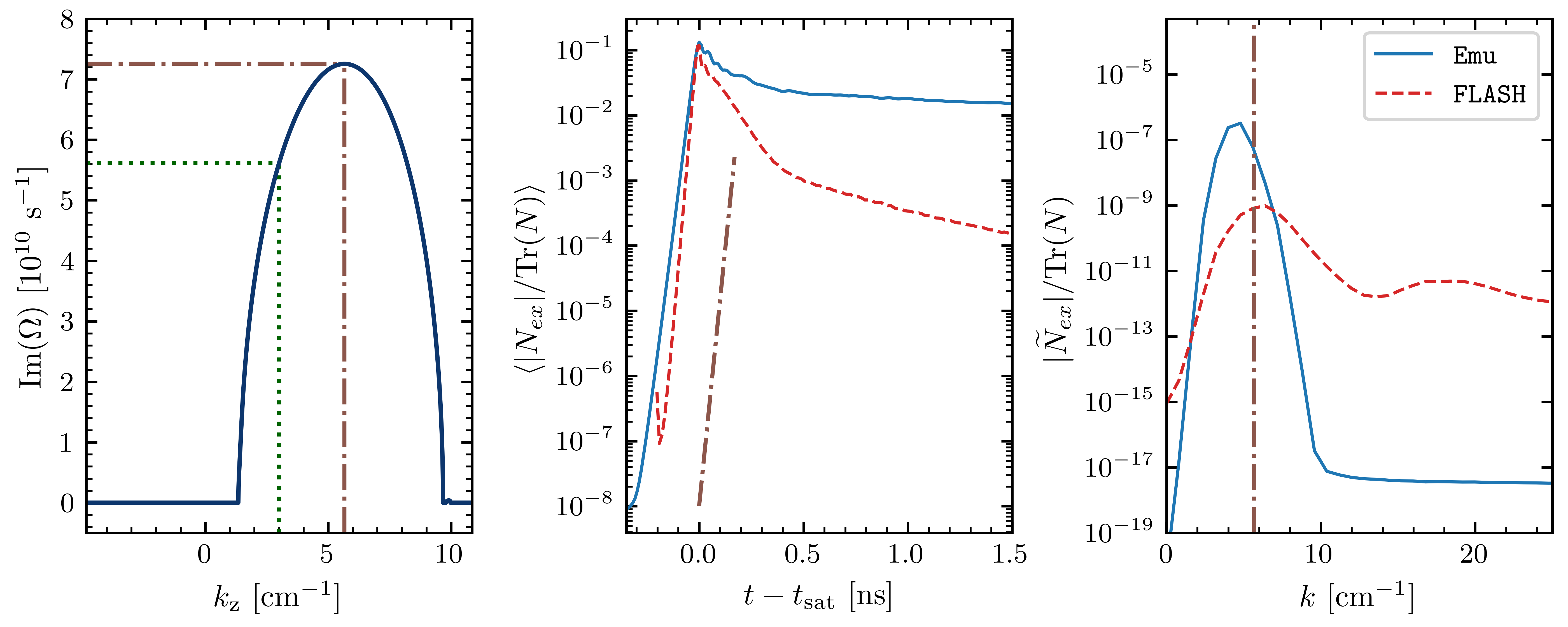}
    \caption{\label{fig:simu_nsm1} Comparison of linear stability analysis predictions and three-dimensional numerical solutions of neutrino flavor transformation (adapted from~\cite{Grohs:2022fyq}) for the NSM 1 point. \emph{Left:} instability growth rate obtained by linear stability analysis for different modes, similarly to Fig.~\ref{fig:LSA_fiducial}. The values corresponding to the zero mode are represented in dotted green lines. \emph{Middle:} evolution of the domain-averaged magnitude of the off-diagonal component of $N$ in \emu and \flash. \emph{Right:} Fourier transform of the off-diagonal component against wavenumber $k$, taken $0.1 \, \mathrm{ns}$ before saturation. Plotting conventions are the same as in Fig.~\ref{fig:simu_fiducial}. The dash-dotted brown line is the linear stability analysis prediction (see Table~\ref{tab:NSM_123_kz}).}
\end{figure*}

\subsubsection{Comparison to numerical simulations}
\label{subsubsec:comparison_num}

Three points from the neutron star merger simulation studied here have been used as initial conditions for a full three-dimensional quantum kinetic simulation of neutrinos in \flash and \emu calculations~\cite{Grohs:2023pgq}, similarly to the test case of Sec.~\ref{sec:fiducial}. They are identified, following the nomenclature from~\cite{Grohs:2023pgq}, as “NSM 1, 2, 3” on Fig.~\ref{fig:slices_NSM}. The predictions of moment LSA and the numerical results from \flash and \emu are gathered in Table~\ref{tab:NSM_123_kz}.

\renewcommand{\arraystretch}{1.4}

\begin{table}[!ht]
    \begin{tabular}{|r|c|c|c|}
    \hline
         & \hphantom{ab}\textsf{LSA}\hphantom{ab} & \hphantom{a}\flash\hphantom{a} & \hphantom{ab}\emu\hphantom{ab} \\ \hline \hline
         \multicolumn{4}{|c|}{“NSM 1” $(X \simeq 20 \, \mathrm{km}, \, Y = 0 \, \mathrm{km}, \, Z \simeq 20 \, \mathrm{km})$}  \\ \hline
        $\quad \Im(\Omega)_\mathrm{max} \ (10^{10} \ \mathrm{s}^{-1})$ & $7.25$  & $8.1$ & $5.6$ \\
        $k_\mathrm{max} \ (\mathrm{cm}^{-1})$ & $5.68$ & $6.4(4)$ & $4.8(4)$ \\ \hline \hline
         \multicolumn{4}{|c|}{“NSM 2” $(X = 0 \, \mathrm{km}, \, Y = 0 \, \mathrm{km}, \, Z \simeq 15 \, \mathrm{km})$}  \\ \hline
        $\quad \Im(\Omega)_\mathrm{max} \ (10^{10} \ \mathrm{s}^{-1})$ &   & $5.2$ & $1.1$ \\
        $k_\mathrm{max} \ (\mathrm{cm}^{-1})$ & & $6.1(4)$ & $3.8(4)$ \\ \hline \hline
         \multicolumn{4}{|c|}{“NSM 3” $(X \simeq 48 \, \mathrm{km}, \, Y = 0 \, \mathrm{km}, \, Z = 0 \, \mathrm{km})$}  \\ \hline
        $\Im(\Omega)_\mathrm{max} \ (10^{10} \ \mathrm{s}^{-1})$ & $2.5$ & $10.7$ & $4.2$ \\
        $k_\mathrm{max} \ (\mathrm{cm}^{-1})$ & $6.6$ & $13.0(5)$ & $6.5(5)$ \\ \hline
    \end{tabular}
    \caption{\label{tab:NSM_123_kz} Linear stability analysis and simulation results for the three points considered in~\cite{Grohs:2023pgq} and identified on Fig.~\ref{fig:slices_NSM}, with $\vec{k}$ restricted to the lepton number flux direction. Here, no unstable mode is found for the NSM 2 point. The uncertainty on $k_\mathrm{max}$ from numerical simulations comes from the limited resolution of the Discrete Fourier Transform with a box size $L = 8 \, \mathrm{cm}$.}
\end{table}

The first point, located at $(X \simeq 25 \, \mathrm{km}, Y = 0 \, \mathrm{km} , Z \simeq 20 \, \mathrm{km})$, was also studied in~\cite{Grohs:2022fyq}. The linear stability analysis results for the growth rate and the wavenumber of the most unstable mode fall between the \flash and \emu ones, and future improvements of the moment method might better the agreement between the two techniques. We represent the linear stability analysis results obtained by sweeping $k_\z$ values along with the numerical results from \emu and \flash calculations in Fig.~\ref{fig:simu_nsm1}. Linear stability analysis can only provide information on the initial exponential growth phase, where similar conclusions to those from the Fiducial test case (see Fig.~\ref{fig:simu_fiducial}) can be drawn. The differences between the \flash and \emu simulations are more significant than in the previous test case, as could be expected given a problem with significantly fewer symmetries (see~\cite{Grohs:2022fyq,Grohs:2023pgq} for initial condition parameters corresponding to those in Table~\ref{tab:fiducial}). Note that when restricting the study to the “zero mode” (see Fig.~\ref{fig:zero_hom_mode}), the instability growth rate is smaller ($\mathrm{Im}(\Omega)_{\mathrm{max},k'=0} \simeq 5.62 \times 10^{10} \, \mathrm{s}^{-1}$). This value can be read on the left panel of Fig.~\ref{fig:simu_nsm1} (dotted green line), the zero mode corresponding here to $k_0 \simeq 3.0 \, \mathrm{cm}^{-1}$. We clearly see in this case that, although unstable, the zero mode is not the \emph{most} unstable mode.

The other two points show more disagreement between the \flash and \emu calculations and with the moment LSA. The NSM 2 point corresponds to a very shallow ELN crossing (see~\cite{Grohs:2023pgq}), for which we find that our analysis does not provide very good results; we come back to this in the next section. In the NSM 3 case, the depth of the ELN crossing is also much smaller than for the NSM 1 point. The length scale of the instability is well captured by the moment LSA, while the growth rate is slightly smaller than \emu and a factor of a few smaller than \flash. While this provides a reasonable estimate for situations where predictions good to a factor of a few are all that is required — which will often be the case given the disparate timescales in neutron star merger simulations — more general applicability of the moment LSA method will require improving its performance in distributions with shallow crossings.

Although the closure itself is a significant limitation to the accuracy of moment methods, our particular choices regarding how to construct a quantum pressure tensor do not seem to have a significant impact on the results. Specifically, choosing different options between Eqs.~\eqref{eq:Pij_thin_1}--\eqref{eq:Pij_thin_2} or \eqref{eq:chi_ab_nonFT}--\eqref{eq:chi_ab} does not alter significantly our results: switching between the four possible combinations shifts $k_\mathrm{max}$ by less than $0.4 \, \%$ (resp. $0.4 \, \%$) and $\Im(\Omega)_\mathrm{max}$ by less than $1.5 \, \%$ (resp. $3 \, \%$) for the NSM 1 (resp. NSM 3) point. None of these choices change the negative determination of instability for the NSM 2 point. This may be a consequence of the fact that throughout the simulation snapshot we have used, most of the neutrinos are close to the optically thick limit. The classically computed Eddington factor is lower than $0.50$ for 99~\% ($\nu_e$), 96~\% ($\nu_x$) and 97~\% ($\bar{\nu}_e$) of the simulation points in our chosen NSM snapshot. This is consistent with the small impact from the different possibilities we are considering for the thin part of the pressure tensor. But while our results appear to be robust, we are reminded that a more accurate “quantum” closure cannot be straightforwardly obtained from a classical one.

\subsection{Electron lepton number crossings}
\label{subsec:ELN_crossing}

The phenomenon of fast flavor instability is known to be related to the presence of ELN crossings (or more precisely, ELN-XLN crossings, see e.g.,~\cite{Richers:2022dqa,Morinaga:2021vmc,Nagakura:2022kic}). Of course, in our situation the initial XLN is always zero since the original NSM simulation groups all heavy lepton neutrinos and antineutrinos together. Therefore, we seek to compare our results to the existence of ELN crossings in the classical neutrino distributions. Intrinsically, a truncated-moment description of the neutrino fluid hides the underlying angular distribution, and only yields information about the distribution properly averaged to give the number and flux densities. 

Various studies have focused on different ways to assess the presence of ELN(-XLN) crossings using angular reconstruction methods \cite{Abbar:2020fcl,Johns:2021taz,Nagakura:2021suv,Richers:2022dqa}, and more recently machine learning procedures~\cite{Abbar:2023kta,Abbar:2023zkm}. We restrict here to the method consistent with the classical simulation that we used to obtain the moments, assuming the same classical maximum entropy moment closure we chose for the stability analysis and that is used to set the initial conditions in \emu~\cite{Johns:2021taz,Richers:2022dqa}.  

\begin{figure}[!ht]
    \centering     
    \includegraphics{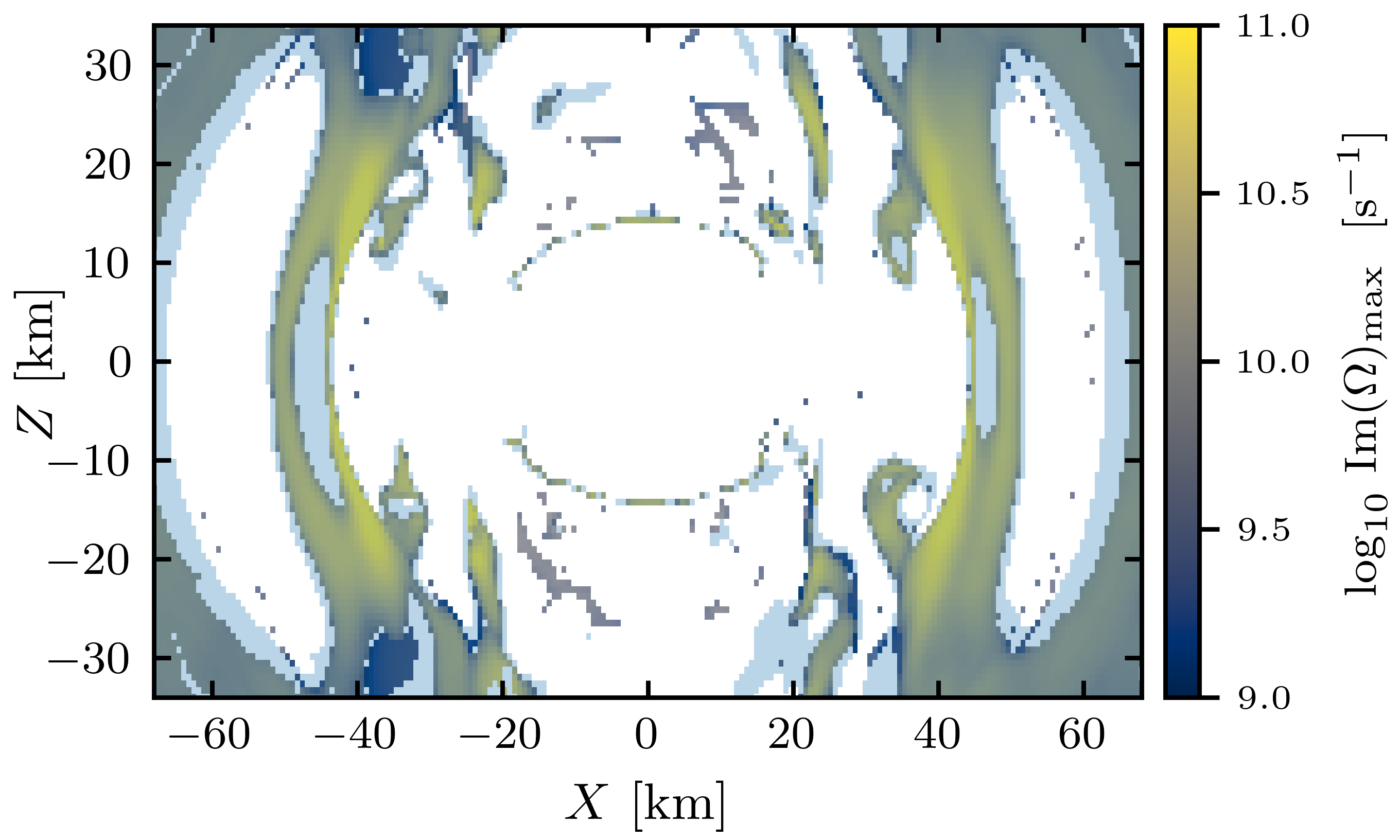}
    \caption{\label{fig:Crossing_slice} On top of the growth rate (cf.~Fig.~\ref{fig:slices_NSM}, left panel), the blue regions indicate where an ELN crossing is determined from Eq.~\eqref{eq:crossing_maxent}.}
\end{figure}

Imposing a maximum entropy (ME) closure amounts to assuming that the underlying neutrino angular distribution maximizes the angular entropy under the constraints of given number density and fluxes. This leads, for the Minerbo closure, to the functional form~\cite{cernohorsky_bludman}
\begin{equation}
\label{eq:dist_ang_ME}
f^\text{ME}_\nu(\vec{x},\vec{\Omega},t) = \frac{N}{4 \pi} \frac{Z}{\sinh(Z)} e^{Z \cos \theta} \ ,
\end{equation}
where $\cos \theta = \vec{\Omega} \cdot \vec{F}/\abs*{\vec{F}}$ is the azimuthal angle with respect to the net number flux. We dropped flavor indices here, implying that this distribution applies to each species separately. The parameter $Z$ satisfies the equation
\begin{equation}
\label{eq:Z_MEC}
    \hat{f} = \mathrm{coth}(Z) - \frac{1}{Z} \ ,
\end{equation}
with $\hat{f}$ the flux factor. ELN crossings can thus be estimated whenever there is a solution of the equation $f_{\nu_e}^\text{ME}(\vec{x},\vec{\Omega},t) = f_{\bar{\nu}_e}^\text{ME}(\vec{x},\vec{\Omega},t)$, i.e. an angle $\theta$ such that
\begin{equation}
\label{eq:crossing_maxent}
     \quad N_{ee} \frac{Z_e}{\sinh(Z_e)}e^{Z_e \cos(\theta - \theta_e)} = \overline{N}_{ee} \frac{\bar{Z}_e}{\sinh(\bar{Z}_e)}e^{\bar{Z}_e \cos(\theta - \bar{\theta}_e)} \ ,
\end{equation}
where $\theta_e$ and $\bar{\theta}_e$ are the respective directions of $\vec{F}_{ee}$ and $\vec{\overline{F}}_{ee}$, taken with respect to a reference direction in the plane $(\vec{F}_{ee},\vec{\overline{F}}_{ee})$ (see~\cite{Richers:2022dqa,Grohs:2023pgq}). The points for which Eq.~\eqref{eq:crossing_maxent} has a solution are depicted on Fig.~\ref{fig:Crossing_slice}, where we compare these a priori unstable regions to the ones found with our moment LSA. Reassuringly, there is generally a good agreement between the two. Although we do not represent it here, the agreement is similarly good across the full three-dimensional NSM simulation.

There are nevertheless some missing regions that can have several origins. We find generally that the moment LSA provides particularly reliable results for deep ELN crossings, but in regions with very shallow crossings our analysis might miss the instability.  However, these regions are at the boundaries between instability and stability, meaning that a propagating neutrino will encounter the FFI at nearly the same place as our current LSA prediction. This should mitigate the overall impact on large-scale simulations of this misalignment of the instability boundary. In addition, our approximate self-consistency procedure (outlined in Appendix~\ref{app:multi_angle_LSA}) can discard some physical modes, though a detailed study of these individual points would circumvent this problem, but at a significant computational cost. Although moment LSA rather under-predicts instability, we find a small number of locations where modes predicted to be unstable by LSA exist but where there is typically an ELN distribution that nearly kisses but does not cross zero. Such disagreements are not unexpected, given the necessary choices that go into the treatment of the pressure tensor and the fact that the current moment LSA method struggles in regions of shallow crossings. 

Importantly, the use of flavor-traced quantities in our semiclassical closure limits the scope of the comparison with classically determined crossings, which allow different flavors to have completely independent distributions. In particular, the distributions compared in~\eqref{eq:crossing_maxent} correctly describe the $\nbnuin_e$ number densities and fluxes, but not the pressure [since we use the flavor-traced Eddington factor~\eqref{eq:chi_ab}, at least for the off-diagonal components, see after Eq.~\eqref{eq:wij_2}]. Linear stability analysis nevertheless provides an efficient framework to test and compare various closure prescriptions beyond the semiclassical expressions~\eqref{eq:Pij_thick}--\eqref{eq:Pij_closed_quantum}, which we leave for future work.

\section{Summary and Conclusions}
\label{sec:conclusion}

One path toward a better assessment and future inclusion of neutrino flavor oscillations in simulations of binary neutron star mergers is to selectively use quantum moments built from the first few classical angular moments often used in such simulations. To understand where and when quantum transport will be necessary, we developed a moment-based linear stability analysis (LSA) method and assessed its efficacy using several metrics.  We first made a comparison between our LSA predictions and both dynamical moment (\flash) and particle-in-cell (\emu) results on a straightforward test case and found good agreement. We then compared results from our moment-based LSA to simulations of flavor transformation in regions extracted from a post-merger general relativistic radiation-hydrodynamics simulation \cite{Foucart:2015gaa,Foucart:2016rxm}.

We confirmed the ubiquity of FFI in NSM simulations by taking a snapshot and applying our LSA to all points. As our method allows us to consider perturbations with any wavevector oriented in any direction, we found that in regions where the FFI should occur, the fastest growing mode is characterized by a wavevector aligned with the direction of the electron lepton number flux ($\vec{F}_{ee} - \vec{\bF}_{ee}$), with values ranging typically up to a few $\mathrm{cm}^{-1}$. We also evaluated the effectiveness of the so-called “zero mode” approximation and found that while it under-estimates the growth rate, it reasonably approximates the spatial regions of instability and often correctly determines the order of magnitude of the instability growth rate. For the simulation region that we worked with, the FFI amplification timescale was generally of order $\sim 0.1 \, \mathrm{ns}$. We also tested our results against regions of ELN angular crossings and found broad consistency, although the edges of the instability regions where the crossings are shallow are not always in precise alignment. We speculate that this could lead to slightly different outcomes in a fully dynamical simulation that would include feedback between flavor transformation and angular crossings.

Our method can be straightforwardly extended to capture additional physics.  By including vacuum and matter terms in the Hamiltonian we can study slow modes and by including collisions, phenomena such as collisional instabilities can be studied. The method can also be extended to multi-energy neutrinos allowing for the study of spectral swaps, splits and the consequences of energy-changing collision terms. 

In this work we have restricted ourselves to semiclassical closures which are minimal extensions of the maximum entropy closure.  While this produced broad agreement in all of our tests, there are subtle details that need improvement. For example, for our choice(s) of closure, there was some ambiguity in how to define the angular distributions with which to determine regions with angular crossings.  In general, there are many such unresolved issues with the closure prescriptions, and in future work, we plan to use our moment LSA method for a more thorough study of closure relationships with an eye toward their inclusion in dynamical simulations. Moment LSA, with its intrinsic computational efficiency, is a particularly attractive tool to begin such a study.

In summary, the overarching goal is to eventually include flavor transformation in situ when performing large-scale simulations of not only merging neutron stars but also core-collapse supernovae and black hole neutron star mergers.  In the short run, since the quasi-steady state reached after fast flavor conversions is not fully understood (see e.g., \cite{Zaizen:2023ihz,Xiong:2023vcm,Abbar:2023ltx} for dedicated studies), phenomenological prescriptions can be used to assess the sensitivity to such neutrino flavor conversions in NSM or CCSN simulations~\cite{Just:2022flt,Li:2021vqj,Ehring:2023abs,Ehring:2023lcd}. Eventually flavor transformation must be incorporated into the simulations via the inclusion of methods that dynamically evolve the neutrinos, e.g.~those of ~\cite{Richers_emu,Grohs:2022fyq}. At present these methods are beyond the reach of computational capabilities and it is necessary to locate a path forward. One possibility is to selectively use quantum approaches only where needed but if this approach is to work, we must be able to identify the appropriate zones in the simulation. The method that we have outlined in this paper is a feasible method to identify such zones. We can now predict FFI in a computationally efficient way by direct use of angular moments of the neutrino field. Our method could be extended to predict other types of instabilities and is also a useful tool for eventually identifying a best-practice quantum closure, a critical problem that must be addressed before quantum kinetic moment methods can be widely adopted.

\section*{Acknowledgments}
We thank Luke Johns for helpful discussions on collisional instabilities. We thank the Institute for Nuclear Theory at the University of Washington for its kind hospitality and stimulating research environment during advanced stages of this work. J.F. is supported by the Network for Neutrinos, Nuclear Astrophysics and Symmetries (N3AS), through the National Science Foundation Physics Frontier Center award No. PHY-2020275.
E.G., J.P.K., and G.C.M. are supported by the Department of Energy Office of Nuclear Physics award DE-FG02-02ER41216. S.R. was supported by a National Science Foundation Astronomy and Astrophysics Postdoctoral Fellowship under award AST-2001760. E.G., J.P.K., and G.C.M. acknowledge support from  U.S. DOE contract Nos. DE-FG0202ER41216 and DE-SC00268442 (ENAF).
F.F. is supported by the Department of Energy, Office of Science, Office of Nuclear Physics, under contract number DE-AC02-05CH11231, by NASA through grant
80NSSC22K0719, and by the National Science Foundation through grant AST-2107932.

\appendix

\section{Moment formalism}
\label{app:moments}

Following~\cite{Shibata:2011kx,Richers:2019grc}, we rewrite the QKEs for the first angular moments of the (anti)neutrino density matrix. To that end, we write the neutrino four-momentum as $p^\alpha = \veps \left(u^\alpha + l^\alpha \right)$, where $l^\alpha$ is a unit normal four-vector ($l^\alpha l_\alpha = 1$) orthogonal to the dimensionless fluid four-velocity $u^\alpha$. We will always ignore the neutrino rest mass by setting the energy equal to the $3$-momentum magnitude $p$. In the comoving frame (which will be our frame of reference), we have $u^\alpha = (1, \vec{0})$ and $l^\alpha = (0, \vec{\Omega})$, such that $\veps = p^0$ is the neutrino energy in this frame. The coordinates are $x^\mu = (t, \vec{x})$.

We assume that we can work with the Minkowski metric in the comoving frame, with metric signature $(-,+,+,+)$. We then define the first angular moments~\cite{Richers:2019grc}: 

\begin{equation}
    \begin{aligned}
        \N_{ab}(x^\mu) &\equiv \int{\dd \veps} \,  \frac{\veps^2}{(2 \pi)^3} \int{\dd{\vec{\Omega}} \, \vrho_{ab}(\veps,\vec{\Omega},x^\mu)} \, , \\
        \F^{\alpha}_{ab}(x^\mu) &\equiv \int{\dd \veps} \, \frac{\veps^2}{(2 \pi)^3}  \int{\dd{\vec{\Omega}} \, l^\alpha \vrho_{ab}(\veps,\vec{\Omega},x^\mu)} \, , \\
        \P^{\alpha \beta}_{ab}(x^\mu) &\equiv  \int{\dd \veps} \, \frac{\veps^2}{(2 \pi)^3}  \int{\dd{\vec{\Omega}} \, l^\alpha l^\beta \vrho_{ab}(\veps,\vec{\Omega},x^\mu)} \, . \\
    \end{aligned}
\end{equation}
These definitions relate to Eq.~\eqref{eq:moments} since, in the comoving frame, 
\begin{equation}
    \N = N \ , \ \ \F^\alpha = (0, \vec{F}) \ \ \text{and} \ \ \P^{\alpha \beta} = \begin{pmatrix}
    0 & \vec{0} \\
    \vec{0}^T & P^{ij}
\end{pmatrix} \, .
\end{equation}
It is readily possible to generalize these expressions to a fully covariant formalism, cf.~\cite{Shibata:2011kx}.

The general moments of the distribution function are then
\begin{multline}
    \M^{\alpha_1 \cdots \alpha_k}(x^\mu) \equiv \int{\dd \veps} \, \frac{\veps^2}{(2 \pi)^3}  \int{\dd{\vec \Omega}  \, \vrho(\veps,\vec \Omega,x^\mu)} \\
    \times (u^{\alpha_1}+l^{\alpha_1})\cdots (u^{\alpha_k}+l^{\alpha_k}) \, .
\end{multline}
Thus, the rank-1 and rank-2 moments read
\begin{align}
    \M^\alpha_{ab} &= \N_{ab} u^\alpha + \F^\alpha_{ab} = \left(N_{ab}, \vec{F}_{ab}\right) \, , \\
    \M^{\alpha \beta}_{ab} &= \N_{ab} u^\alpha u^\beta + \F^\alpha_{ab}u^\beta + \F^{\beta}_{ab}u^\alpha + \P^{\alpha \beta}_{ab} \nonumber \\ &= \begin{pmatrix} N_{ab} & \vec{F}_{ab} \\
    \vec{F}_{ab}^T & P^{ij}_{ab}
    \end{pmatrix} \, .
\end{align}
Note that all previous expressions are defined accordingly for antineutrinos.

In Minkowski spacetime, the moment QKEs are written~\cite{Shibata:2011kx}
\begin{equation}
\label{eq:moment_QKE_rel}
    \partial_\alpha \M^{\alpha \beta} = \S^\beta \, ,
\end{equation} 
with $\S^\beta$ the source term, which reads for the refractive part
\begin{equation}
    \S^\beta_{ab} = - \i \, \int{\dd \veps} \,  \frac{\veps^2}{(2 \pi)^3} \int{\dd \vec{\Omega} \, \frac{p^\beta}{\veps}  \left [H,\vrho \right]_{ab}} \, .
\end{equation}
It is straightforward to show that the self-interaction Hamiltonian~\eqref{eq:H_self} can be written
\begin{equation}
H_{\nu \nu} = - \sqrt{2} G_F \frac{p^\alpha}{\veps} \left( \M_\alpha - \overline{\M}_\alpha^* \right) \, .
\end{equation}
Therefore the QKE~\eqref{eq:moment_QKE_rel} becomes:
\begin{equation}
\label{eq:moment_QKE_rel_fin}
 \partial_\alpha \M^{\alpha \beta} = - \i \left[ \mathcal{H}_\alpha, \M^{\alpha \beta}\right] \, ,
\end{equation}
with $\mathcal{H}_\alpha = - \sqrt{2} G_F \left( \M_\alpha - \overline{\M}_\alpha^* \right)$. Components $\beta = 0$ and $\beta = j$ of Eq.~\eqref{eq:moment_QKE_rel_fin} give respectively equations~\eqref{eq:QKE_moment_N} and~\eqref{eq:QKE_moment_F}.

\section{Multi-angle linear stability analysis}
\label{app:multi_angle_LSA}

In order to determine whether the predictions from moment linear stability analysis are self-consistent, we use a simplified multi-angle analysis to assess how well moments should be able to capture the predicted instability.

Namely, from the eigenvector $(A_{ex}, B_{ex}^i, \bA_{xe}, \bB_{xe}^i)^T$ corresponding to the fastest growing mode, we can deduce the associated angular distribution of eigenvectors in a multi-angle analysis. If this distribution is very narrow in angular space, one cannot expect a two-moment method to accurately describe such a mode, thus failing the self-consistency test.

As this procedure is solely used as an additional consistency check, we restrict the analysis to distributions of neutrinos that maintain azimuthal symmetry. For simplicity, we assume this symmetry is around the $\z$-axis, which generally corresponds to the direction of the fastest growing mode $\vec{k}_\mathrm{max}$ (when $\vec{k}_\mathrm{max} \nparallel \vec{z}$ as in Fig.~\ref{fig:slice_3D}, the azimuthal symmetry is taken around $\vec{k}_\mathrm{max}$). The polar angles $\theta_n$ are discretized, and we write $\mu_n \equiv \cos \theta_n$. The self-interaction Hamiltonian~\eqref{eq:H_self} for a momentum $\vec{p}$ pointing in the direction $\theta_n$ can be written:
\begin{equation}
H_{\nu \nu, n} = \sqrt{2} G_F \sum_{m}{(1-\mu_n \mu_m) \left[N_m - \bN_m^* \right]} \, ,
\end{equation}
where $N_m$ is the number density of neutrinos in directional bin $m$ such that $\sum_m N_m=N$. That is,
\begin{equation}
    N_{ab,m}(t,\vec{x}) = \Delta \Omega_m {\int{\dd p}} \, \frac{p^2}{(2 \pi)^3} \vrho_{ab}(t, \vec{x},\vec{p}_m) \, ,
\end{equation}
with $\vec{p}_m$ of magnitude $p$ in the direction $\vec{\Omega}_m$. Note that, with the assumption of azimuthal symmetry, $\Delta \Omega_m = 2 \pi \Delta \mu_m$. The QKEs in direction $\mu_n$ then read:
\begin{equation}
    \label{eq:QKE_multiangle}
\begin{aligned}
\i \left(\partial_t + \vec{v}_n \cdot \vec{\nabla} \right) N_n &= [H_{\nu \nu, n}, N_n] \, , \\
\i \left(\partial_t + \vec{v}_n \cdot \vec{\nabla} \right) \bN_n &= -[H_{\nu \nu, n}^*, \bN_n] \, ,
\end{aligned}
\end{equation}
with $\vec{v}_n = \vec{p}_n/p$ the velocity (of norm $1$ in natural units), pointing in the direction $\theta_n$. We linearly perturb the density matrices similarly to~\eqref{eq:perturb_N}, such that:
\begin{equation}
\delta N_n = \begin{pmatrix} 0 & a_{e x,n} \\ a_{x e, n} & 0 \end{pmatrix} e^{- \i(\Omega t - \vec{k} \cdot \vec{x})} + \mathrm{h.c.} \, ,
\end{equation}
and likewise for $\delta \bN_n$. Therefore, the self-interaction Hamiltonian is also perturbed at first order by:
\begin{equation}
\delta H_{\nu \nu,n} = \sqrt{2} G_F \sum_{m}{(1-\mu_n \mu_m) \left[\delta N_m - \delta \bN_m^*\right]} \, .
\end{equation}
The linearized QKEs~\eqref{eq:QKE_multiangle} read 
\begin{align}
\i (\partial_t + \vec{v}_n \cdot \vec{\nabla})\delta N_n &= [H^{(0)}_{\nu \nu, n}, \delta  N_n] + [\delta H_{\nu \nu, n}, N^{(0)}_n] \, , \\
\i (\partial_t + \vec{v}_n \cdot \vec{\nabla})\delta \bN_n &= -[H^{(0)*}_{\nu \nu, n}, \delta  \bN_n] - [\delta H_{\nu \nu, n}^*, {\bN}^{(0)}_n] \, .
\end{align}
Collecting the terms proportional to $e^{-\i (\Omega t - \vec{k} \cdot \vec{x})}$, we get:

\begin{widetext}
\begin{align}
    \left( \Omega - \mu_n k_\z \right) a_{ex, n} &= \sqrt{2} G_F \sum_{m}{(1-\mu_n \mu_m) \left[N_{ee,m}-N_{xx,m} - \bN_{ee,m} + \bN_{x x, m}\right]} \times a_{ex,n} \nonumber \\
    &\qquad \qquad \qquad \qquad - \sqrt{2} G_F \sum_{m}{(1-\mu_n \mu_m)\left[a_{ex,m} - \bar{a}_{xe,m}\right]} \times (N_{ee, n} - N_{x x, n}) \, , \\
    \left( \Omega - \mu_n k_\z \right) \bar{a}_{xe, n} &= \sqrt{2} G_F \sum_{m}{(1-\mu_n \mu_m)\left[N_{ee,m}-N_{xx,m} - \bN_{ee,m} + \bN_{x x, m}\right]} \times \bar{a}_{xe,n} \nonumber \\
    &\qquad \qquad \qquad \qquad - \sqrt{2} G_F \sum_{m}{(1-\mu_n \mu_m)\left[a_{ex,m} - \bar{a}_{xe,m}\right]} \times (\bN_{ee, n} - \bN_{x x, n}) \, ,
\end{align}
that is
\begin{multline}
(\Omega - \mu_n k_\z) a_{ex, n}  =  \sqrt{2} G_F \left(N_{ee} - N_{xx} - \bN_{ee} + \bN_{xx}\right) a_{ex, n} - \mu_n \sqrt{2} G_F \left(F_{ee}^\z - F_{xx}^\z - \bF_{ee}^\z + \bF_{xx}^\z \right) a_{ex, n} \\
- \sqrt{2} G_F  (N_{ee, n} - N_{x x, n}) \left[ \left(A_{ex} - \bA_{xe}\right) - \mu_n \left(B_{ex}^\z - \bB_{xe}^\z\right)\right]  \, ,
\end{multline}
and likewise for antineutrinos.
\end{widetext}

These equations can be rewritten, using the quantities $\Omega'$ and $\vec{k}'$ introduced in Eq.~\eqref{eq:def_omp_kp},
\begin{align}
\Omega' a_{ex, n} &=  \mu_n k'_\z a_{ex, n} 
 - \sqrt{2} G_F (N_{ee, n} - N_{x x, n}) \nonumber \\ &\qquad \qquad \times \left[ \left(A_{ex} - \bA_{xe}\right) - \mu_n \left(B_{ex}^\z - \bB_{xe}^\z\right)\right]  \, , \label{eq:LSA_multiangle_nu} \\
 \Omega' \bar{a}_{xe, n} &=  \mu_n k'_\z \bar{a}_{xe, n} 
 - \sqrt{2} G_F (\bN_{ee, n} - \bN_{x x, n}) \nonumber \\ &\qquad \qquad \times \left[ \left(A_{ex} - \bA_{xe}\right) - \mu_n \left(B_{ex}^\z - \bB_{xe}^\z\right)\right] \, . \label{eq:LSA_multiangle_nubar}
\end{align}
If we sum~\eqref{eq:LSA_multiangle_nu} and \eqref{eq:LSA_multiangle_nubar} over $n$, we logically find back the moment equations~\eqref{eq:QKE_lin_A} and \eqref{eq:QKE_lin_Abar} (with $\vec{k}' = k'_\z \vec{u}_\z$ because of the azimuthal symmetry around $\z$ assumption). In order to use this formalism to “check” the moment LSA results, we express the multi-angle eigenvector elements as a function of the moment eigenvector elements:
\begin{align}
    \frac{a_{ex,n}}{N_{ee,n} - N_{x x, n}} &= - \sqrt{2} G_F \frac{A_{ex} - \bA_{xe} - \mu_n (B_{ex}^\z - \bB_{xe}^\z)}{\Omega' - \mu_n k'_\z} \, , \label{eq:multi-angle_consist1} \\
    \frac{\bar{a}_{xe,n}}{\bN_{ee,n} - \bN_{x x, n}} &= - \sqrt{2} G_F \frac{A_{ex} - \bA_{xe} - \mu_n (B_{ex}^\z - \bB_{xe}^\z)}{\Omega' - \mu_n k'_\z} \, . \label{eq:multi-angle_consist2}
\end{align}
In a multi-angle LSA, we do not introduce the moments ($A_{ex},B_{ex}^\z,\dots$) but write the matrix equation involving only the perturbations $\{a_{ex,n},\bar{a}_{xe,n}\}$ and look for unstable modes. Note that, if we sum~\eqref{eq:multi-angle_consist1} and \eqref{eq:multi-angle_consist2} over $n$, we can rewrite these two equations in the matrix form:
\begin{equation}
    \begin{pmatrix} I_0 + 1 & - I_1 \\ -I_1 & I_2 - 1 \end{pmatrix} \begin{pmatrix} A_{ex} - \bA_{xe} \\ B_{ex}^\z - \bB_{xe}^\z \end{pmatrix} = 0 \, ,
\end{equation}
where
\begin{equation}\label{eq:moment_multi_angle}
I_\ell \equiv \sqrt{2} G_F \sum_{n}{\frac{(N_{ee,n} - N_{xx,n}) - (\bN_{ee,n}-\bN_{xx,n})}{\Omega' - \mu_n k'_\z}\mu_n^\ell} \, .
\end{equation}
Setting the determinant to zero then gives the equation usually solved to get the dispersion relation in azimuthal symmetry (e.g., \cite{Dasgupta:2016dbv,Johns:2019izj}):
\begin{equation}
(I_0 +1)(I_2 -1) - I_1^2 = 0 \, .
\end{equation}

In an ideal moment analysis, should the closure reflect perfectly all the truncated higher-order moments, the results should agree with the multi-angle LSA. Here, we solely use the previous equations as a consistency check of our moment LSA results: given the moment eigenvectors $\{A_{ex},\bA_{xe},B_{ex}^\z,\bB_{xe}^\z\}$, Eqs.~\eqref{eq:multi-angle_consist1} and~\eqref{eq:multi-angle_consist2} provide a relationship between $\mu_n$ and $a_{ex,n}$, i.e. more generally the function $a_{ex}(\theta)$, which represents the distribution in angular space of the unstable part of the density matrix. The diagonal coefficients $\nbNin_{aa,n}$ are given by the angular distribution associated to the Minerbo closure~\eqref{eq:dist_ang_ME}, but azimuthally averaged so that $\theta_n$ is the polar angle with respect to the direction of $\vec{k}'_\mathrm{max}$. More precisely, the Minerbo closure angular distribution is axially symmetric around the direction of the species' flux, but in this simplified multi-angle analysis we instead assume such a symmetry around the direction of $\vec{k'}$. We thus need to define the “effective” distribution in direction $\theta$ with respect to $\vec{k'}$:
\begin{equation}
\label{eq:effective_rho}
    \hat{\vrho}(\theta) = \frac{1}{2 \pi} \int_{0}^{2 \pi}{\dd{\varphi} \, f_{\nu}^\mathrm{ME}(\cos(\vec{\Omega}_{\theta,\varphi},\vec{F}))} \, , 
\end{equation}
where the vector $\vec{\Omega}_{\theta,\varphi}$ is defined by its spherical coordinates $(\theta, \varphi)$ around $\vec{k'}$. This effectively azimuthally symmetric distribution around $\vec{k}'$ is such that $2\pi  \hat{\vrho}(\theta)$ is the total number of neutrinos having a momentum making an angle $\theta$ with $\vec{k}'$. We compute Eq.~\eqref{eq:effective_rho} numerically using the Python function \texttt{quad} for each value of $\theta_n$.

\begin{figure}[!ht]
    \centering  \includegraphics{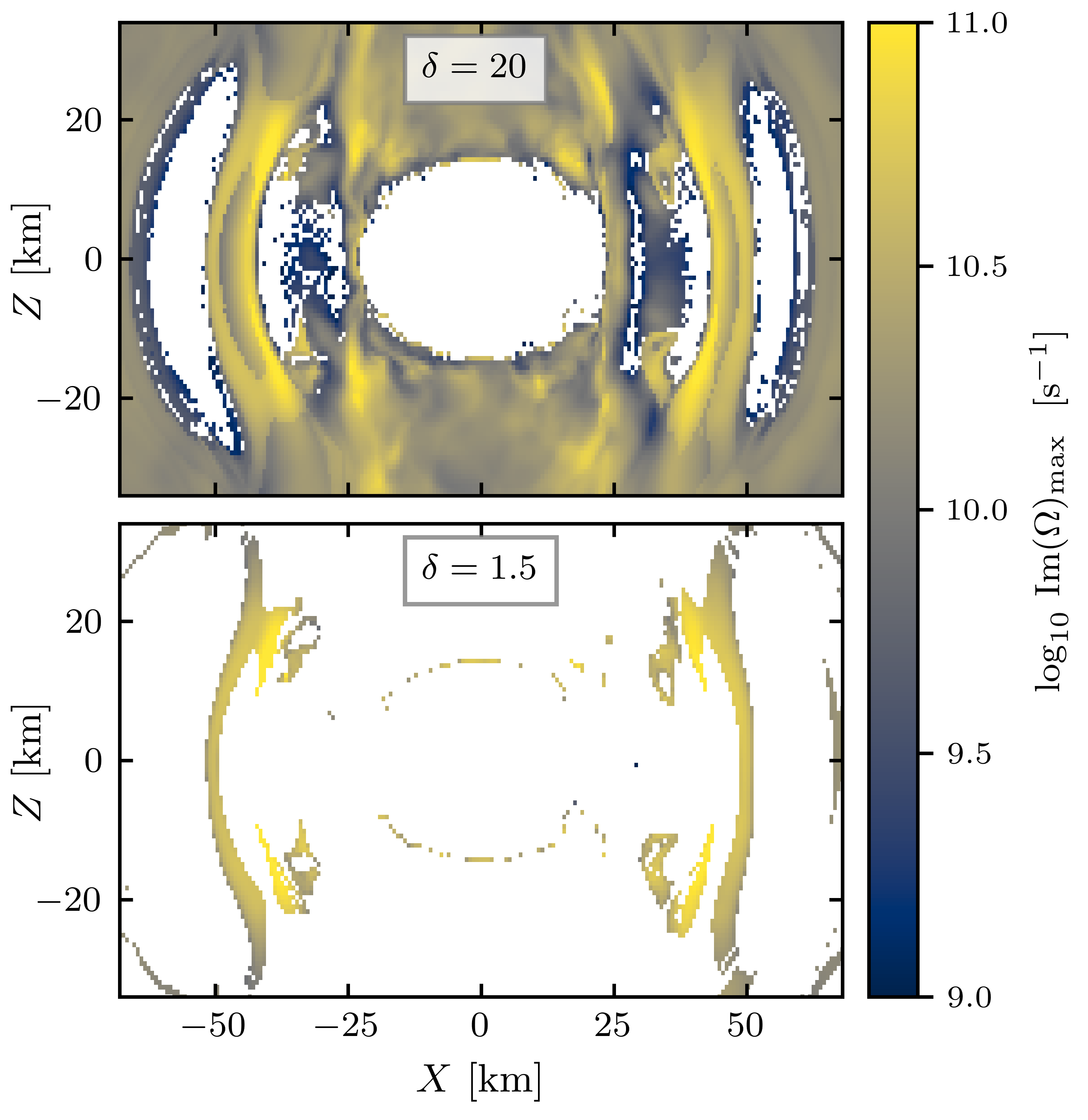}
    \caption{\label{fig:cutoff} Fast flavor instability growth rate for the NSM slice $\{Y = 0 \, \mathrm{km}\}$, for a 3D search in $\vec{k}$ space, varying the cutoff of the consistency test. The large value ($\delta = 20$) overestimates the possibility of FFI, notably in the polar regions (cf. Fig.~\ref{fig:Crossing_slice} where the regions with an ELN crossing are represented), while the low value ($\delta = 1.5$) rejects too many points.}
\end{figure}

It is illusory to expect our two-moment method to accurately describe a situation where the distribution $a_{ex}(\theta)$ is narrowly peaked around a given direction, as any small angular features would only be captured by higher-order moments, or equivalently by a “perfect” closure. Therefore, should the results from moment LSA correspond to such an ill-suited distribution $a_{ex}(\theta)$, we should conservatively discard the associated mode. In concrete terms, if $a_{ex}(\theta)$ is forward-peaked, there exists a range of $\theta$ where $a_{ex}$ sharply drops off, i.e., a range of $\theta$ where $|\dd a_{ex}/\dd \mu|$ is large (with $\mu = \cos \theta$). We thus impose a cutoff $\delta$ on the derivative of $\abs{a_{ex}(\mu)}$. That is, if the multi-angle LSA for a given value of $\vec{k'}$ gives a distribution commensurate with the criterion
\begin{equation}
\label{eq:cutoff}
    \frac{1}{\mathrm{max}_\mu \{ \abs{a_{ex}(\mu)} \}} \abs{\frac{\dd \abs{a_{ex}}}{\dd \mu}} > \delta \, ,
\end{equation}
we reject that particular mode as inconsistent with the moment LSA. The smaller the value of $\delta$, the more conservative we are in selecting reasonable unstable modes. Of course, by setting an arbitrary cutoff we can miss some physical modes, but we carried out some tests to determine an optimum value for $\delta$. Throughout this paper, we have taken $\delta = 4$. As an example, we show in Fig.~\ref{fig:cutoff} the results for the same slice as Fig.~\ref{fig:slice_3D}, with a too large value of $\delta$ (which does not exclude some “fictitious” modes) and a too small value (which largely underestimates the occurrence of FFI).

\section{Collisional instabilities}
\label{app:collisional_inst}

Fast flavor instabilities are not the only flavor conversion channel that can take place in an environment such as a neutron star merger. In particular, the possibility of collisional instabilities that are associated with an asymmetry between the neutrino and antineutrino interaction rates was first evidenced in~\cite{Johns:2021qby} and subsequently studied in, e.g., \cite{Padilla-Gay:2022wck,Johns:2022yqy,Lin:2022dek,Xiong:2022vsy,Xiong:2022zqz,Liu:2023pjw,Liu:2023vtz,Akaho:2023brj,Fiorillo:2023ajs}. These works confirmed that collisional instabilities may appear in the neutrino decoupling regions of supernovae and mergers. In this appendix, we use the simple criterion given in~\cite{Johns:2021qby} to assess the possibility of collisional instabilities in our NSM simulation. A detailed study of the role of collisions in this system is beyond the scope of this work, but might be warranted in the future given the possible interplay between fast flavor and collisional instabilities~\cite{Johns:2022yqy,Xiong:2022zqz}. We solely aim here at giving an indication on the potential regions of interest for collision-induced flavor instabilities and where our results should therefore be taken more cautiously.

In a relaxation-time (or “damping”) approximation, collisions are described by a total rate $\Gamma$ for neutrinos and $\bar{\Gamma}$ for antineutrinos (which can be separated in absorption/emission and scattering processes). Namely, these rates are the averages of the rates for individual species, that is $\Gamma = (\Gamma_{\nu_e}+\Gamma_{\nu_x})/2$ and likewise for $\bar{\Gamma}$. For distributions that are purely flavor-diagonal (which is our starting point from the NSM simulation), the collisional instability criterion given in Ref.~\cite{Johns:2021qby} reads
\begin{equation}
\label{eq:coll_instab_criterion}
    \abs{(\Gamma - \bar{\Gamma})(N_{ee}+\bN_{ee}-2 N_{xx})} \gtrsim \abs{(\Gamma + \bar{\Gamma})(N_{ee}-\bN_{ee})} \, ,
\end{equation}

\noindent where we assumed that $N_{xx} = \bN_{xx}$, consistent with the NSM simulation we look at. The growth rate of the collisional instability is then (see Eq.~(14) in \cite{Johns:2021qby}):
\begin{equation}
\label{eq:coll_instab_growth}
    \mathrm{Im}(\Omega)_\mathrm{coll} \simeq \pm \frac{\Gamma - \bar{\Gamma}}{2} \abs{\frac{N_{ee} + \bN_{ee} - 2 N_{xx}}{N_{ee} - \bN_{ee}}} - \frac{\Gamma + \bar{\Gamma}}{2} \, .
\end{equation}
Note that these expressions assume homogeneity and isotropy, and can therefore only give an indication of where collisional instabilities could take place in an actual NSM.

\begin{figure}[!ht]
    \centering    
    \includegraphics{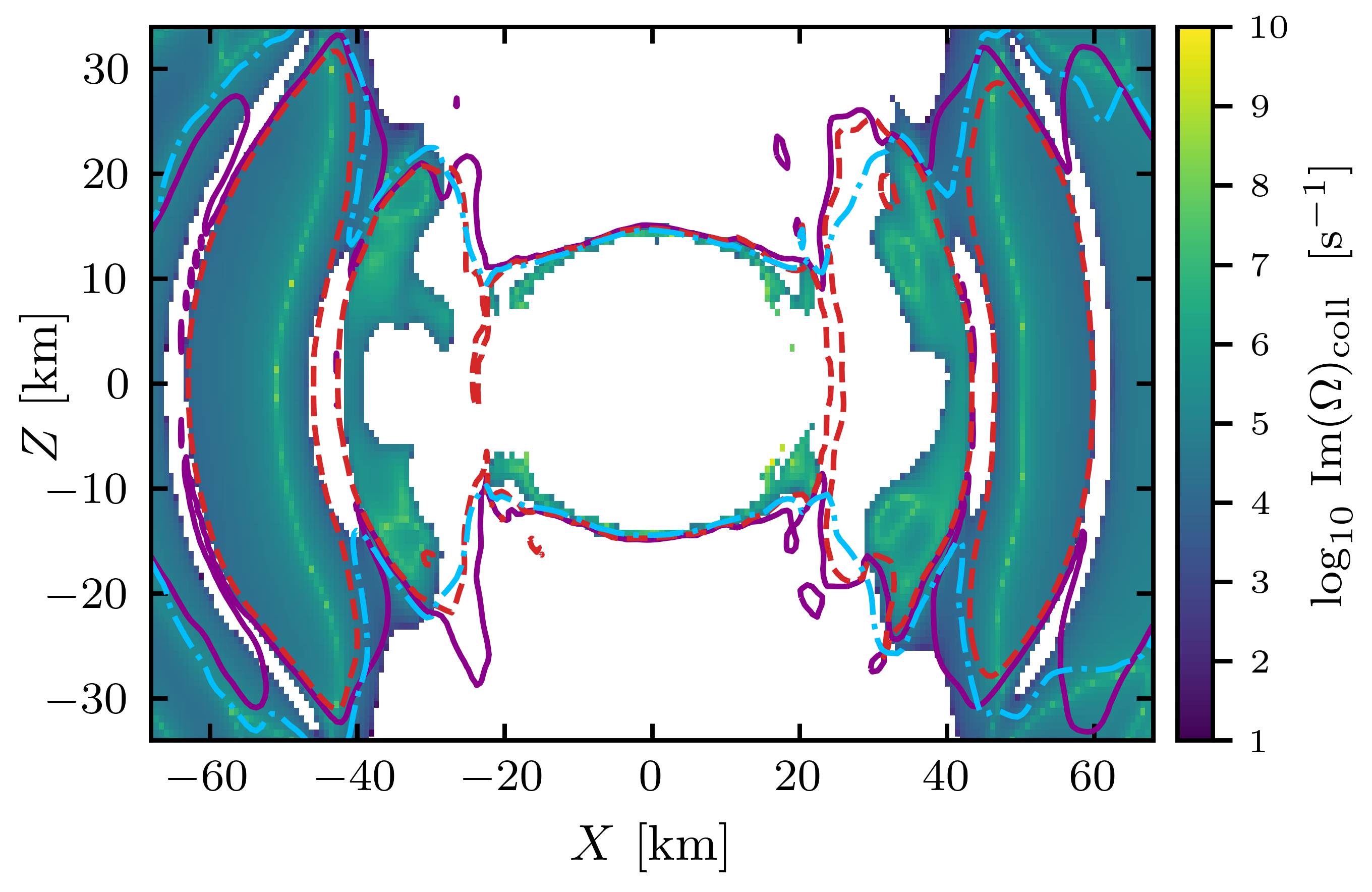}
    \caption{\label{fig:coll_instab} Collisional instability growth rate [Eq.~\eqref{eq:coll_instab_growth}], on the same slice as the left panel of~Fig.~\ref{fig:slices_NSM}. It is generally several orders of magnitude smaller than the FFI growth rate. For reference, the $\hat{f} = 0.2$ flux factor contours, which indicate the “neutrinospheres”, are represented for $\nu_e$ (solid purple line), $\bar{\nu}_e$ (dashed red line) and $\nu_x$ (dash-dotted blue line).}
\end{figure}

\begin{figure}[!ht]
    \centering  
    \includegraphics[width=0.9\columnwidth]{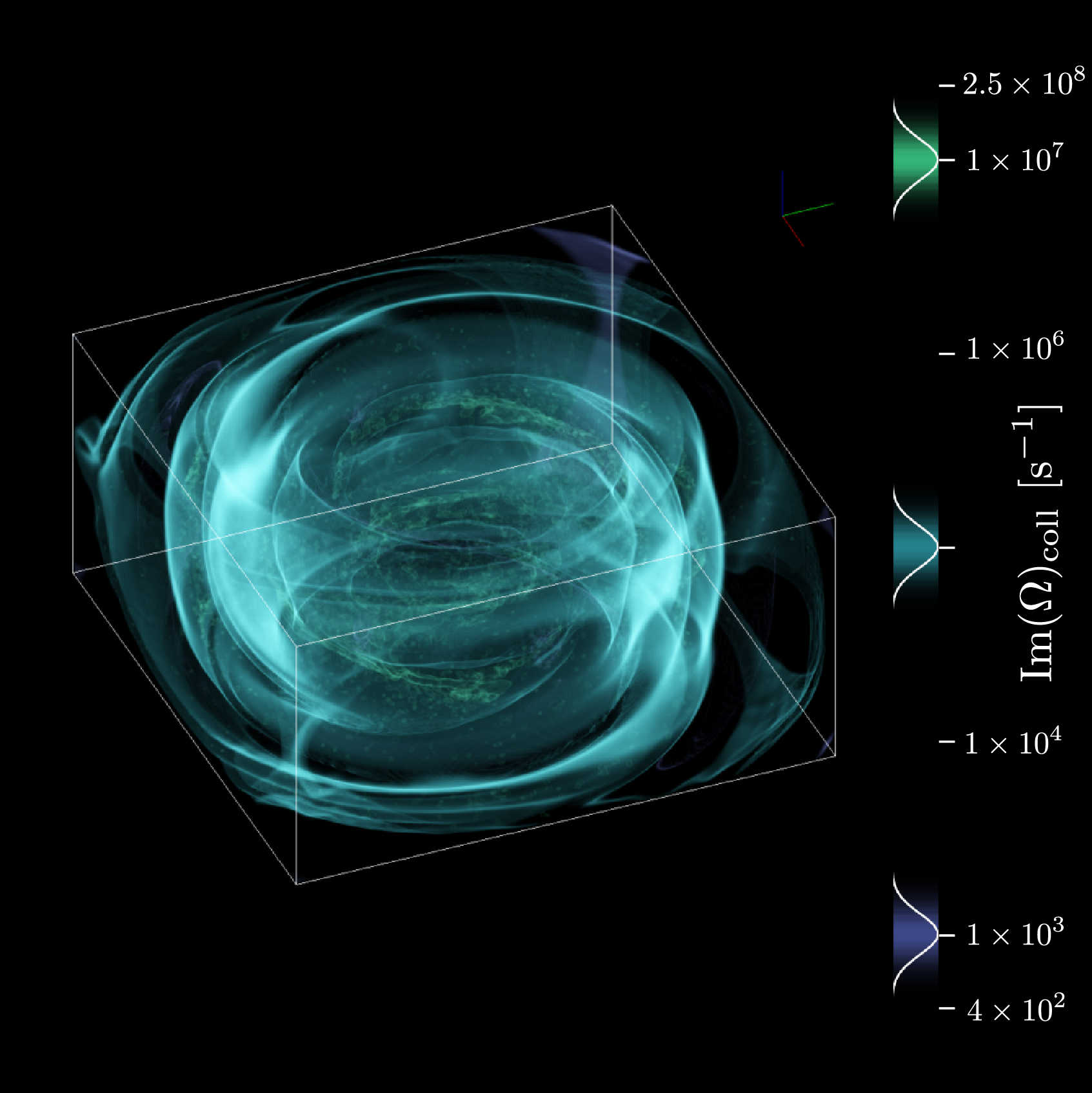}
    \caption{\label{fig:coll_instab_vol} Volume rendering of the collisional instability growth rate across the NSM simulation, predicted using Eq.~\eqref{eq:coll_instab_growth}. Three colored contours are centered respectively around the growth rate values $10^3 \, \mathrm{s^{-1}}$, $10^{5} \, \mathrm{s^{-1}}$, and $10^7 \, \mathrm{s^{-1}}$.}
\end{figure}

We compute the collision rates from the neutrino interaction library \texttt{NuLib},\footnote{\texttt{NuLib} is available at \href{http://www.nulib.org}{http://www.nulib.org}.} using the temperature, matter density, average neutrino energy and electron fraction from the simulation snapshot~\cite{Foucart:2015gaa,Foucart:2016rxm} and assuming the SFHo equation of state~\cite{Steiner:2012rk}. The instability growth rate, at any point where the criterion~\eqref{eq:coll_instab_criterion} is verified, is shown on Fig.~\ref{fig:coll_instab} for the slice $\{Y = 0 \, \mathrm{km}\}$ and on Fig.~\ref{fig:coll_instab_vol} in a volume rendering across the simulation. 

As can be seen by comparing the instability regions with the $\hat{f}=0.2$ flux factor contours drawn on Fig.~\ref{fig:coll_instab}, the regions of collisional instability are rather well delimited by the boundary between optically-thick and optically-thin regions. This is a posteriori consistent with the assumption of isotropy that goes into Eqs.~\eqref{eq:coll_instab_criterion} and \eqref{eq:coll_instab_growth}, although going beyond this limiting assumption will be unavoidable in the future.

\medbreak

We estimate that collisional instabilities could take place in vast regions of the post-merger simulation, but should be hidden by FFI which are orders of magnitude faster, except in regions which are stable to FFI but appear to host a collisional instability. In particular, including the collision term in the QKEs in those regions will be a necessary improvement to confidently evaluate the possible flavor transformation mechanisms.

\bibliography{references}

\end{document}